\documentclass[useAMS,usenatbib]{mn2e}
\usepackage{graphicx}
\usepackage{lscape}
\usepackage{url} 
\usepackage[dvips]{color}
\usepackage{array}
\usepackage{aas_macros}
\usepackage{upgreek}
\usepackage[T1]{fontenc}
\usepackage{times}
\usepackage{amsmath}
\usepackage{aecompl} 
\usepackage{multicol}
\usepackage{multirow}
\usepackage{mathtools}
\usepackage{nicefrac} 
\DeclarePairedDelimiter\abs{\lvert}{\rvert}

\newcommand{\PROCESSED}{50~per cent} 
\newcommand{\NEWPSR}{60} 
\newcommand{\REDETECTIONu}{912} 
\newcommand{\REDETECTIONp}{435} 
\newcommand{\MISSED}{22} 
\newcommand{\REDETECTIONmsp}{eight} 
\newcommand{\MSPyield}{ten} 
\newcommand{\NEWSOLUTIONiso}{30} 
\newcommand{\NEWSOLUTIONbin}{three}
\newcommand{\HasPdot}{32} 
\newcommand{\NEWNOSOLUTION}{28} 
\newcommand{\SIDELOBE}{198} 
\newcommand{\INBEAM}{709} 
\newcommand{\CALIB}{54} 
\newcommand{\PPDOT}{32} 

\def\gtrsim{\mathrel{\hbox{\rlap{\hbox{\lower3pt\hbox{$\sim$}}}\hbox{\raise2pt\hbox{$>$}}}}}

\newcolumntype{R}[1]{>{\raggedleft\let\newline\\\arraybackslash\hspace{0pt}}p{#1}}
\newcolumntype{x}[1]{>{\centering\let\newline\\\arraybackslash\hspace{0pt}}p{#1}}
\title[HTRU$-$XII. The Galactic plane survey]
  {The High Time Resolution Universe Pulsar Survey XII : Galactic plane acceleration search and the discovery of \NEWPSR{} pulsars}
\author[C. Ng et al.]
  {C.~Ng$^1$, D.~J.~Champion$^1$, M.~Bailes$^{2,3}$, E.~D.~Barr$^{2,3}$, S.~D.~Bates$^{4,5}$, N.~D.~R.~Bhat$^{2,3,6}$,
  \newauthor
  M.~Burgay$^7$, S.~Burke-Spolaor$^8$, C.~M.~L.~Flynn$^{2,3}$, A.~Jameson$^2$, S.~Johnston$^9$, M.~J.~Keith$^5$, 
  \newauthor
  M.~Kramer$^{1,5}$, L.~Levin$^{10}$, E.~Petroff$^{2,3,9}$, A.~Possenti$^7$, B.~W.~Stappers$^5$,
  \newauthor
  W.~van~Straten$^{2,3}$, C.~Tiburzi$^{7,11}$, R.~P.~Eatough$^1$, A.~G.~Lyne$^5$\\
  $^1$Max-Planck-Institut f\"{u}r Radioastronomie, 
      Auf dem H\"{u}gel 69, D-53121 Bonn, Germany \\
  $^2$Centre for Astrophysics and Supercomputing, 
      Swinburne University of Technology, Mail H39, 
      PO Box 218, VIC 3122, Australia \\
  $^3$ARC Centre of Excellence for All-Sky Astronomy (CAASTRO), 
      Mail H30, Swinburne University of Technology, 
      PO Box 218, Hawthorn, VIC 3122, Australia \\
  $^4$National Radio Astronomy Observatory, 
      PO Box 2, Green Bank, WV 24944, USA \\
  $^5$Jodrell Bank Centre for Astrophysics,
      University of Manchester, Alan Turing Building,
      Oxford Road, Manchester M13 9PL, United Kingdom \\
  $^6$International Centre for Radio Astronomy Research,
      Curtin University, Bentley, WA 6102, Australia \\
  $^7$INAF-Osservatorio Astronomico di Cagliari,
      via della Scienza 5, 09047 Selargius, Italy\\
  $^8$NASA Jet Propulsion Laboratory, M/S 138-307, 
      Pasadena CA 91106, USA \\
  $^9$CSIRO Astronomy $\&$ Space Science, 
      Australia Telescope National Facility, 
      PO Box 76, Epping, NSW 1710, Australia \\  
  $^{10}$Department of Physics, West Virginia University, 
      Morgantown, WV 26506, USA \\
  $^{11}$Dipartimento di Fisica, Universit$\grave{a}$ di 
      Cagliari, Cittadella Universitaria 09042 
      Monserrato (CA), Italy}
\date{Accepted 2015 April 2. Received 2015 March 27; in original form 2014 November 26}

\pagerange{\pageref{firstpage}--\pageref{lastpage}} \pubyear{2015}

\def\LaTeX{L\kern-.36em\raise.3ex\hbox{a}\kern-.15em
    T\kern-.1667em\lower.7ex\hbox{E}\kern-.125emX}

\begin{document}

\label{firstpage}

\maketitle

\begin{abstract}
\noindent
We present initial results from the low-latitude Galactic plane region of the High Time Resolution Universe pulsar survey conducted at the Parkes 64-m radio telescope. We discuss the computational challenges arising from the processing of the terabyte-sized survey data. Two new radio interference mitigation techniques are introduced, as well as a partially-coherent segmented acceleration search algorithm which aims to increase our chances of discovering highly-relativistic short-orbit binary systems, covering a parameter space including potential pulsar-black hole binaries. We show that under a constant acceleration approximation, a ratio of data length over orbital period of $\approx$ 0.1 results in the highest effectiveness for this search algorithm. From the \PROCESSED{} of data processed thus far, we have re-detected \REDETECTIONp{} previously known pulsars and discovered a further \NEWPSR{} pulsars, two of which are fast-spinning pulsars with periods less than 30\,ms. PSR~J1101$-$6424 is a millisecond pulsar whose heavy white dwarf (WD) companion and short spin period of 5.1\,ms indicate a rare example of full-recycling via Case A Roche lobe overflow. PSR~J1757$-$27 appears to be an isolated recycled pulsar with a relatively long spin period of 17\,ms. In addition, PSR~J1244$-$6359 is a mildly-recycled binary system with a heavy WD companion, PSR~J1755$-$25 has a significant orbital eccentricity of 0.09, and PSR~J1759$-$24 is likely to be a long-orbit eclipsing binary with orbital period of the order of tens of years. Comparison of our newly-discovered pulsar sample to the known population suggests that they belong to an older population. Furthermore, we demonstrate that our current pulsar detection yield is as expected from population synthesis.
\end{abstract}


\section{INTRODUCTION} \label{sec:intro}
Pulsars are extraordinary natural laboratories with extremely high density and gravity impossible to re-create on Earth, hence they provide exclusive insights to a rich variety of fundamental physics and astronomy. To discover more pulsars we have performed the High Time Resolution Universe (HTRU) survey: a blind survey of the southern sky with the 64-m Parkes radio telescope in Australia \citep{HTRU1} and a twin survey of the northern sky with the 100-m Effelsberg radio telescope in Germany \citep[the HTRU-North survey;][]{Barr2013}. The HTRU survey uses multi-beam receivers and backends constructed with state-of-the-art technology, providing unprecedented high time and frequency resolution, allowing us to probe deeper into the Galaxy than previous efforts with these telescopes. We split the survey into three regions of the sky, namely the low-latitude Galactic plane survey, the medium-latitude survey, and the high-latitude all sky survey, each tailored to achieve specific science goals.

The low-latitude Galactic plane region is where the most relativistic binaries are expected to be found \citep{Belczynski2002}. Pulsars in tight binaries orbiting other compact objects, for example neutron stars and, potentially, black holes, are of great interest as their strong gravitational fields provide the best tests of General Relativity \citep[GR;][]{Wex2014} and other theories of gravity \citep{Freire2012}. The best example of such a binary system so far is the double pulsar system \citep{Burgay2003,Lyne2004}. The double pulsar has been used to obtain four independent tests of GR and GR has passed the most stringent test, regarding the shape of the Shapiro delay, with a measurement uncertainty of only 0.05~per cent \citep{Kramer2006b}. The number and the precision of GR tests increase as the binary systems to be discovered become more relativistic. Hence, one of the main aims of the HTRU Galactic plane survey is precisely the discovery and study of ultra-compact relativistic binary systems in short orbits.

In this region of sky within Galactic latitude $\pm$3.5$^{\circ}$, we employ the longest HTRU integrations of 72\,min per pointing to maximise our sensitivity. This means that the HTRU low-latitude Galactic plane data set will be capable of revealing many pulsars that were not luminous enough to be detected by previous surveys. These pulsars might increase the sample of sources that glitch, which could lead to improved knowledge of the interior of neutron stars \citep{Espinoza2011}. They will give us useful insights into the interstellar medium via the study of their dispersion and rotation measures to reveal a picture of the magnetic field of the Milky Way \citep[see e.g.,][]{Noutsos2008}. They also provide us with an important handle on the lower end of the luminosity distribution function of the Galactic plane pulsar population, valuable knowledge for the planning of survey strategies for the next generation of radio telescopes, such as MeerKAT\footnote{http://www.ska.ac.za/meerkat/index.php} and the Square Kilometre Array\footnote{https://www.skatelescope.org/} (SKA). Furthermore, the archive of the HTRU Galactic plane survey will continue to produce science through future data re-examination. The long dwell time of these observations is also favourable for the detection of transient and nulling sources deep within the Galactic plane \citep[see e.g., related work by][]{Petroff2014}. 

However, the sheer volume of the high-resolution HTRU Galactic plane data set poses great challenges in data manipulation and analysis. Normally, long integration length ($t_{\rm{int}}$) provides an increase in sensitivity. Nonetheless, this is not exactly the case when it comes to searching for tight-orbit relativistic pulsar binaries: a periodicity search in the Fourier domain is the standard method employed in most pulsar surveys. However, the high orbital acceleration attained by fast relativistic binaries results in a Doppler shift in the spin frequency of the pulsar as a function of the orbital phase. The pulsar signal is thus smeared across neighbouring spectral bins of the Fourier power spectrum, hence a reduction in the sensitivity of the periodicity search. Furthermore, the width of the Fourier spectral bin is defined by $1/t_{\rm{int}}$. As a result, the longer the integration time, the larger the portion of the orbit we cover in a particular observation, but also the narrower the Fourier spectral bin. These two effects combined lead to more severe consequences for Fourier spectral smearing. For these reasons, the search algorithm of the Galactic plane survey presented in this paper requires significant modifications over that used in the medium- and high-latitude survey \citep{HTRU1}. Various search techniques targeting binary pulsars exist, which compensate for, or in some cases fully recover, the reduction in detectability due to orbital motion \citep[see e.g.,][]{Handbook2004}. Here we present an innovative segmented search technique which aims to increase our chances of discoveries of highly-accelerated relativistic short-orbit binary systems, including potential pulsar-black hole binaries. We stress that the depth of the parameter space to which the survey data can be explored is highly dependent on the available computing resources. Optimisation of pulsar searching algorithms is thus crucial in the era of data intensive astronomy, and the computational challenges faced by the HTRU Galactic plane survey will be applicable for the planning of the SKA.

In this paper we focus on the data processing and initial results of the HTRU Galactic plane survey. In Section~\ref{sec:HTRUsystem} we describe the survey strategy and the observation set-up. In Section~\ref{sec:Processing} we present the data processing algorithm with emphasis on the newly incorporated elements, namely two radio-frequency interference (RFI) mitigation schemes (Section~\ref{sec:RFI}) and an innovative `partially-coherent segmented acceleration search' technique (Section~\ref{sec:AccSearch}). We report on the re-detections of known pulsars in Section~\ref{sec:known} and the discoveries of {\NEWPSR}{} pulsars in Section~\ref{sec:new}. We highlight the individual pulsars of interest in Section~\ref{sec:individuals} and we compare the newly-discovered pulsars with the known pulsar population in Section~\ref{sec:population}. In Section~\ref{sec:yield} we compare our discovery rate with the estimated survey yield and in Section~\ref{sec:conclusion} we present our conclusions.

\section{GALACTIC-PLANE SURVEY STRATEGY} \label{sec:HTRUsystem}
\begin{table}
\centering
 \caption{Characteristic minimum detectable flux density ($S_{\rm{1400,min}}$) for the HTRU Galactic plane survey. Considering normal pulsars and MSPs as two separate groups, we note the minimum, mean, and maximum of the duty cycle $\delta$ of these two groups respectively, and we derive the respective $S_{\rm{1400,min}}$. }
\begin{tabular}{lllll}
  \hline
   & \multicolumn{2}{c}{MSPs} &  \multicolumn{2}{c}{Normal pulsars}  \\
   & $\delta$ ($\%$) & $S_{\rm{1400,min}}$ (mJy) & $\delta$ ($\%$) & $S_{\rm{1400,min}}$ (mJy) \\
  \hline
  min & 0.28 & 0.013 & 0.014 & 0.0030 \\
  mean & 11.54 & 0.092 & 4.21 & 0.053 \\
  max & 65.31 & 0.35 & 57.29 & 0.29 \\
  \hline \label{tab:Smin}
 \end{tabular}
\end{table}

The Galactic plane survey covers a strip along the inner Galactic plane, with the central beam of all scheduled pointings between Galactic longitude $-80\degr<\textit{l}<30\degr$ and latitude $|b|$ $<3.5\degr$ (see Fig.~\ref{fig:skyprocessed}). The observational set-up is the same as that described in \citet{HTRU1}. To summarise, the HTRU survey observations were made using the 20-cm multibeam receiver \citep{Multibeam1996} on the 64-m Parkes radio telescope. This receiver is designed for efficient sky surveying, allowing simultaneous observations with its 13 receiver beams each separated by $\sim$30$'$ (Fig.~\ref{fig:grid}). Each receiver beam has a full width at half-maximum (FWHM) of 14$'$.4, and thus requires four interleaving pointings to cover an area in mosaic style (see Fig.~\ref{fig:skyprocessed}). A bandwidth of 400\,MHz is recorded for each beam, but the introduction of low-pass hardware filters against RFI generated by the Thuraya~3 geostationary communications satellite means that we have a final usable bandwidth of 340\,MHz centred at 1352\,MHz.  

We employ the Berkeley-Parkes-Swinburne Recorder (BPSR) digital backend and we sample 1024 channels across the observing bandwidth giving an intra-channel frequency resolution of 0.39\,MHz, at a sampling rate of $t_{\rm{samp}}=64\,\upmu$s. These are significant improvements from the previously most successful pulsar survey, the Parkes Multibeam Pulsar Survey \citep[PMPS;][]{Manchester2001}, where the frequency resolution was 3\,MHz and time resolution 250\,$\upmu$s. These mean that the HTRU survey has high frequency resolution to help the removal of interstellar dispersion and is sensitive to the transient sky down to 64\,$\upmu$s. In addition, the observations were decimated to 2\,bits per sample. This is another advantage compared to PMPS where its 1-bit digitisation might have reduced sensitivity for weak signals and bright radio bursts. Based on these survey specifications, we can use the radiometer equation (see Equation~(\ref{eq:SNExp})) to calculate the characteristic minimum detectable flux density at 1.4\,GHz ($S_{\rm{1400,min}}$) for the HTRU Galactic plane survey. This quantity depends on pulsar duty cycle ($\delta$), hence we have derived the characteristic $S_{\rm{1400,min}}$ with respect to the minimum, mean and maximum $\delta$ of all published pulsar data from the ATNF Pulsar Catalogue\footnote{http://www.atnf.csiro.au/people/pulsar/psrcat/} \citep[\textsc{psrcat};][]{PSRCAT}. We list in Table~\ref{tab:Smin} the corresponding $S_{\rm{1400,min}}$ distinguishing between normal pulsars and separately for millisecond pulsars\footnote{\protect\citet{Lee2012} derived an empirical definition to classify MSP. For simplicity, we adopt a definition of $P \le 30$\,ms for MSP throughout this paper.} (MSPs).

Observations for the HTRU Galactic plane survey at Parkes took place between November 2008 and December 2013, which comprise 1230 scheduled pointings each with 72-min long observations with just over $2^{26}$ samples. The observations have their two polarizations summed, before being written to tape for storage. Several corrupted pointings (due to severe RFI contamination or hardware issues) were re-observed, hence finally 1246 pointings were recorded, resulting in 263\,terabytes of observational raw data.  

\section{PROCESSING} \label{sec:Processing}
Two processing pipelines are applied to the data. Our initial, `standard search' pipeline, also known as the \textsc{hitrun} processing pipeline, is outlined in \citet{HTRU1}. This algorithm follows the typical procedures of pulsar searching such as that described in \citet{Handbook2004}. Firstly, spurious signals in the data, such as those created by radio frequency interference (RFI), are identified and excised. Next, the observation is dedispersed to compensate for the frequency-dependent delay caused by the free electrons along the line of sight. As the amount of dispersion is dependent on the \textit{a priori} unknown distance of the pulsar to be discovered, we trial a wide range of potential dispersion measures (DMs) between 0 and 3000\,cm$^{-3}$\,pc, which sums to a total of 1069 DM trials per data set. Each of the dedispersed time series is then Fourier transformed. We sum the second, fourth, eighth and the 16th harmonic Fourier spectra respectively \citep[see `incoherent harmonic summing' technique in][]{Taylor1969} and identify significant signals created by any pulsating sources. Based on the false-alarm probability \citep{Handbook2004}, this survey has a signal-to-noise (S/N) threshold of $\sim9$. We have nonetheless inspected every potential pulsar candidate with a S/N above 8 by eye. A second pipeline, the `partially-coherent segmented acceleration search', aims to improve the detectability of binary pulsars and is detailed in Section~\ref{sec:AccSearch}. In addition, advancements regarding RFI mitigation are implemented and are presented in Section~\ref{sec:RFI}. A single-pulse related analysis \citep[see e.g.,][]{Spolaor2011} has not yet been carried out and will be conducted in future data re-processing.

\begin{figure*}
\centering
\setlength\fboxsep{0pt}
\setlength\fboxrule{0pt}
\fbox{\includegraphics[width=16cm]{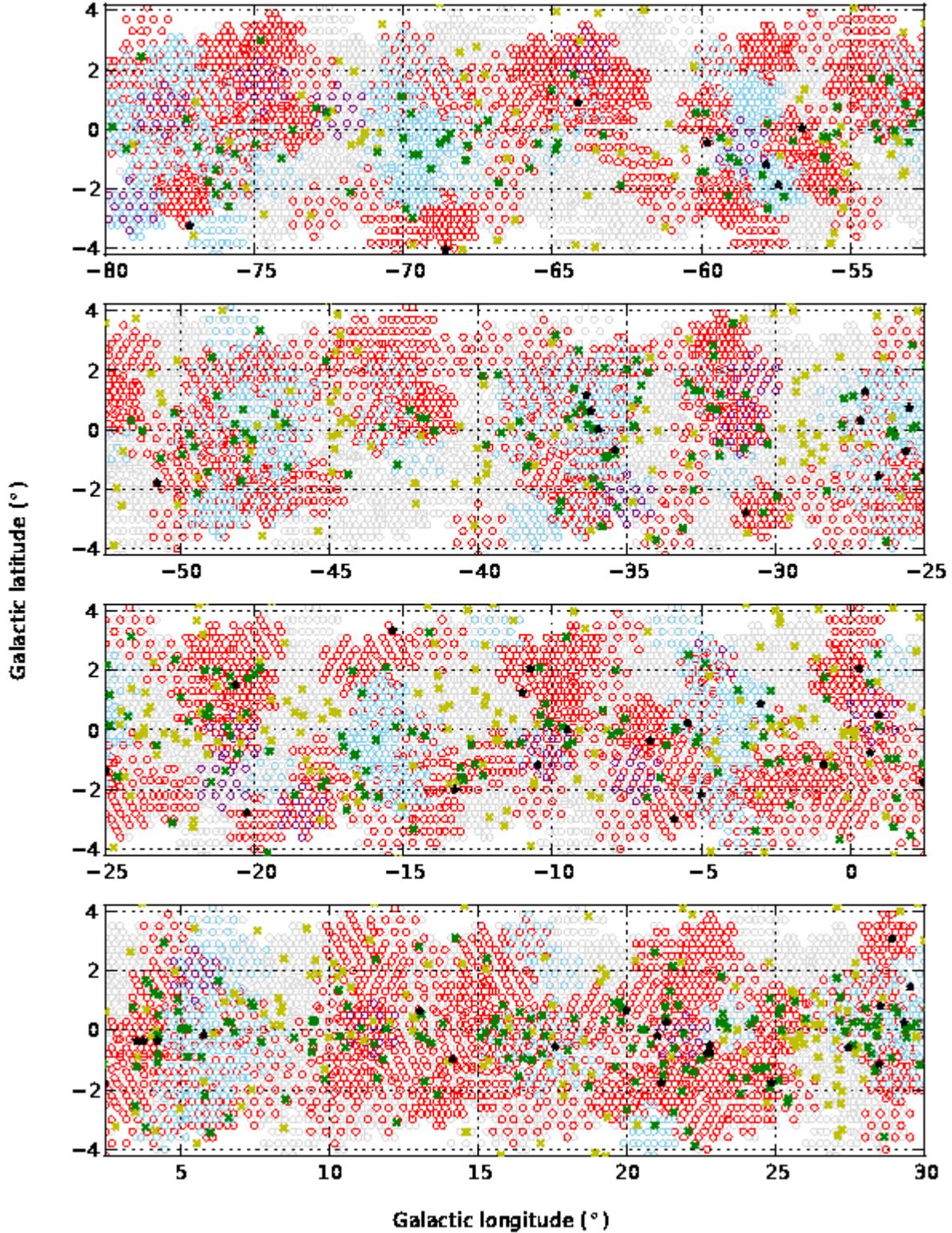}}
\caption{The spatial distribution of the processed observations from the HTRU Galactic plane survey. Grey circles denote the 1230 scheduled pointings for the complete Galactic plane survey. Blue circles denote pointings which have been processed with the `standard search' pipeline (amounting to $\sim$14~per cent of all pointings), whereas red circles are pointings which have been processed with the `partially-coherent segmented acceleration search' pipeline (amounting to $\sim$35~per cent of all pointings). A small portion of 156 beams of observations have been processed with both pipelines to check for compatibility between the two algorithms and are represented by purple circles. Yellow crosses show the locations of previously known pulsars and green crosses show the \REDETECTIONp{} re-detected pulsars. The black stars show the \NEWPSR{} pulsars discovered thus far. }
\label{fig:skyprocessed}
\end{figure*}

To date, 615 pointings of the Galactic plane survey have been processed, which is \PROCESSED{} of the survey. The spatial distribution of the processed data can be seen in Fig.~\ref{fig:skyprocessed}. These processed pointings are not contiguous and only depend on availability of data at the location of the computing facilities. 

\subsection{RFI mitigation} \label{sec:RFI}
\citet{HTRU1} described two RFI removal procedures employed as part of the \textsc{hitrun} pipeline: the removal of RFI-affected spectral channels in the frequency domain targeting narrow channel interference, and the replacement of time samples contaminated by impulsive RFI with noise generated from random sampling of the uncontaminated surrounding data. Two extensions have been incorporated in the Galactic plane survey, both exploiting the fact that RFI is terrestrial hence usually non-dispersed (i.e., most prominent at $\rm{DM}=0$\,cm$^{-3}$\,pc) and often appears in multiple receiver beams, in contrast to celestial sources which are point-like and tend to show up in only one beam, unless they are very bright. 

\subsubsection{Time domain} \label{sec:RFIt}
\begin{figure}
\centering
\includegraphics[width=2.9in]{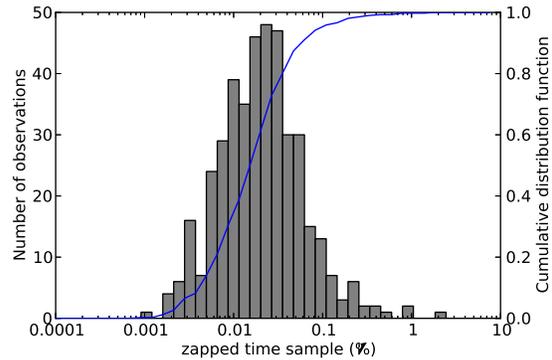}
\caption{Histogram showing the statistics of the percentage of time samples removed per observation, as a result of the automatically generated RFI mask. The blue line shows the corresponding cumulative distribution function.} (Statistics taken from 414 observations, i.e. 34~per cent of the survey)
\label{fig:timezap}
\end{figure}
Since November 2010 an automated scheme to generate empirical time-domain RFI masks has been incorporated into the BPSR backend. This RFI mitigation algorithm has been developed from the work of \citet{Kocz2012}. During each HTRU survey observation, non-dispersed (i.e., at $\rm{DM}=0$\,cm$^{-3}$\,pc) time series are output every 10\,s from BPSR. Each of these 10-s time segments are auto- and then cross-correlated to form a covariance matrix and are subsequently decomposed into eigenvalues. We apply a threshold to identify RFI-affected time samples, namely a cut at 6$\sigma$ for any signal that appears in more than four beams and a lower cut at 4$\sigma$ if the signal is present in all 13 beams. These potentially contaminated time samples are recorded to a log file, which is stored together with the un-corrected observation. The RFI mask is applied to all beams of a single pointing during subsequent off-line data processing to replace the bad time samples with random noise. Fig.~\ref{fig:timezap} shows the statistics of the percentage of masked time samples per observation for the HTRU Galactic plane survey, which also serves as a measure of the quality of the survey data. Typically, for each observation, $\sim0.03$~per cent of the time samples are flagged as RFI-affected, and no observation has $>2$~per cent of masked time samples. For the observations taken before this implementation, we apply the time domain RFI mitigation technique as described in Section~4.1.1 of \citet{HTRU1}.

\subsubsection{Fourier domain} \label{sec:RFIfourier}
\begin{figure*}
\centering
\setlength\fboxsep{0pt}
\setlength\fboxrule{0pt}
\fbox{\includegraphics[width=16cm]{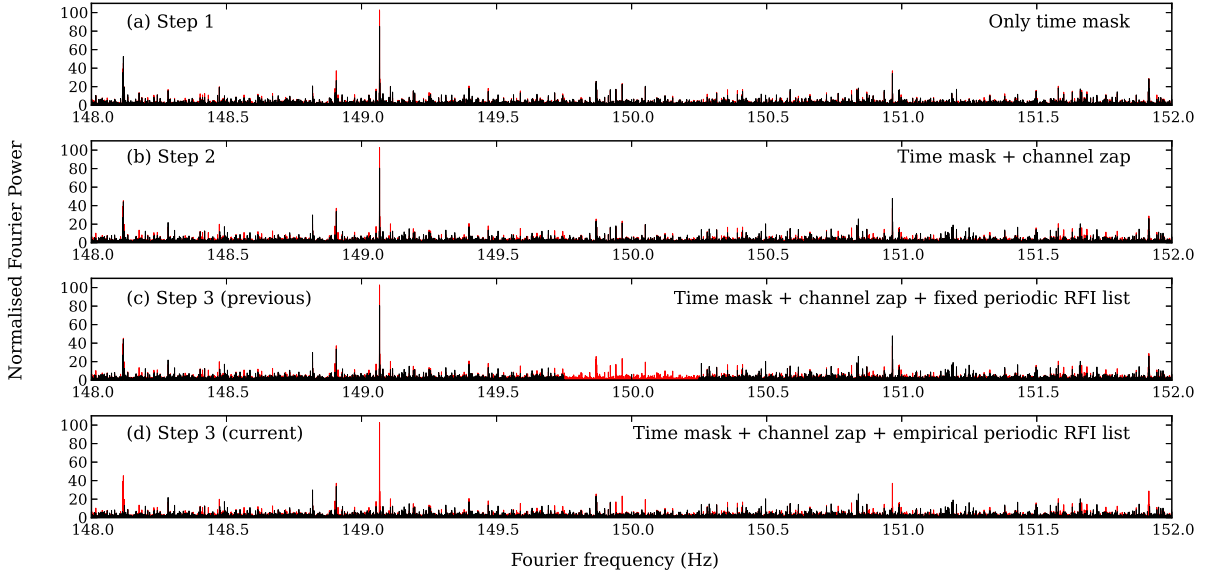}}
\caption{Comparison of the effectiveness in mitigating periodic RFI by applying varying extents of RFI cleaning. We show the power spectrum created from the time series at $\rm{DM}=0$\,cm$^{-3}$\,pc of an actual survey observation, zooming into the region around 150\,Hz where RFI due to harmonics of the Australian mains power supply can be seen clearly. In each panel, the original uncleaned spectrum is plotted in red, whereas the spectrum after each RFI cleaning procedure is plotted in black. In this particular case, the empirically-generated periodic RFI list removed only 0.01~per cent of the Fourier spectral bins (panel~(d)), but is still more effective than applying the fixed periodic RFI list which always removed 0.57~per cent of the spectrum (panel~(c)). We note that subtle differences in the spectra are a result of the change in weighting due to normalisation.} \label{fig:RFI-4stages}
\end{figure*}

The presence of any periodic RFI is most readily identified in the Fourier domain. Nonetheless, great caution must be taken when flagging periodic RFI, to prevent accidentally masking a genuine pulsar signal which is also periodic. As part of the \textsc{hitrun} pipeline, fixed periodic RFI lists have been applied to mask known RFI periodicities, such as that of the Australian mains power supply at $\sim$50\,Hz and its harmonics. However, the RFI environment is temporally varying and a fixed list is far from ideal. We have investigated a more robust method of empirically identifying birdie periodicities. For each observation, we create the power spectrum of each of the 13 beams from their time series at $\rm{DM}=0$\,cm$^{-3}$\,pc. A power threshold ($\mathcal{P}_{\rm{thres}}$) based on false-alarm probability can be calculated with the following equation from \citet{Handbook2004},
\begin{equation}
\mathcal{P}_{\rm{thres}} = - \ln \left( \frac{1}{2n_{\rm{samp}}}\right)\,,
\end{equation}
where $n_{\rm{samp}}$ is the number of time samples used to create the power spectrum and for this survey is 2$^{26}$, which gives a $\mathcal{P}_{\rm{thres}} \approx 19$. We then compare the power spectrum Fourier bin by Fourier bin, identifying Fourier frequencies which exceed $\mathcal{P}_{\rm{thres}}$ in more than four beams and flagging them as periodic RFI. Under this scheme, typically $<0.01$~per cent of the Fourier spectrum is removed for each data set. This compares very favourably to the fixed periodic RFI list incorporated as part of the \textsc{hitrun} pipeline, which contains Fourier frequencies associated with known RFI and always results in the removal of 0.57~per cent of the Fourier spectrum. Note that for the acceleration search algorithm as described in Section~\ref{sec:AccSearch}, as the time re-sampling at every acceleration trial would shift any RFI periodicities in the time series, one periodic RFI list per acceleration trial has to be created. 

Fig.~\ref{fig:RFI-4stages} compares the effectiveness in mitigating periodic RFI by applying varying stages of the above-mentioned techniques on one of the survey observations. In all cases, the original uncleaned spectrum is plotted in red, and the spikes corresponding to harmonics of the Australian mains power supply at $\sim$150\,Hz can be clearly seen. Plotted in black are the spectra after each RFI cleaning procedure, and in theory no spike should remain, as there is no pulsar in this observation. In panel~(a) only the time domain mask as described in Section~\ref{sec:RFIt} has been applied, which might have been effective for removing impulsive RFI but remains insensitive in the Fourier space. In panel~(b) also the frequency channels with excessive power are removed as described in Section~4.1.1 of \citet{HTRU1}, but the result is almost identical to the previous panel. In panel~(c) we apply the fixed periodic RFI list incorporated as part of the \textsc{hitrun} pipeline, which turns out to be masking a part of the spectrum that is relatively RFI-free, but is not able to identify some of the neighbouring narrow spikes. Finally, panel~(d) shows the result of applying the empirically-determined multi-beam Fourier domain RFI mitigation method presented here. Occasionally a few RFI-related spikes (for example those at $\sim$148.8 and 148.9\,Hz), although appearing significant by eye inspection, are still missed out by this RFI mitigation technique, as they are present in less than our chosen conservative threshold of four beams. Nonetheless, most of the prominent spikes have been successfully identified and masked. Indeed, this technique has enabled several of the pulsar discoveries presented here, which from our retrospective checks show that they would not have been found otherwise.

\subsection{Acceleration search} \label{sec:AccSearch}
The challenges related to the searches for binary pulsars have been mentioned in Section~\ref{sec:intro}. To fully account for the effects of unknown orbits on pulsars in our data is an almost impossible task, as theoretically one should search five of the Keplerian orbital parameters\footnote{Out of the six Keperian parameters, the longitude of ascending node is related to the 3D orientation of the binary orbit, which is not relevant in pulsar searching. Hence only the reminding five parameters are considered.} as well as all possible DMs and spin periods, resulting in a large 7-dimensional parameter space which would be extremely computationally expensive. 

A more manageable strategy is to approximate any unknown orbital motion as a simple line-of-sight constant acceleration, i.e., $v(t)=at$, combined with a `time domain resampling' technique \citep[see e.g.,][]{Johnston1991}. By quadratically stretching or compressing a time series by the amount dictated by a particular acceleration, the time series is re-binned into equal time steps in the rest frame of an inertial observer with respect to the pulsar in binary orbit. This resampled time series can then be Fourier transformed to coherently search for peaks in the power spectrum just like the standard periodicity search. This `time domain resampling' technique has been a frequent choice for previous pulsar surveys targeting binary pulsars. Notable examples are the 47~Tucanae observations by \citet{Camilo2000} and the PMPS re-analysis by \citet{Eatough2013}. In the following we detail the implementation of such an acceleration search for the HTRU Galactic plane survey.

\subsubsection{The ratio of data length over orbital period, $r_{\rm{orb}}$} \label{sec:onetenth}
The constant acceleration approximation is equivalent to the best-fitting tangent to a quadratic $v(t)$ curve, and its effectiveness thus depends on the ratio of the integration length ($t_{\rm{int}}$) to the orbital period of the pulsar ($P_{\rm{orb}}$). Here we define this ratio to be $r_{\rm{orb}}$, where
\begin{equation}
r_{\rm{orb}} = \frac{t_{\rm{int}}}{P_{\rm{orb}}} = \frac{t_{\rm{samp}} \times n_{\rm{FFT}}}{ P_{\rm{orb}}}\,.
\label{eq:rorb}
\end{equation}
Note that $t_{\rm{int}}$ is the product of the time sampling rate, $t_{\rm{samp}}$, and the number of samples used in the \textit{Fast Fourier Transform} (FFT), $n_{\rm{FFT}}$. This number should be a power of two for maximum computational efficiency of the FFTs. 

\begin{figure}
\centering
\includegraphics[width=3.1in]{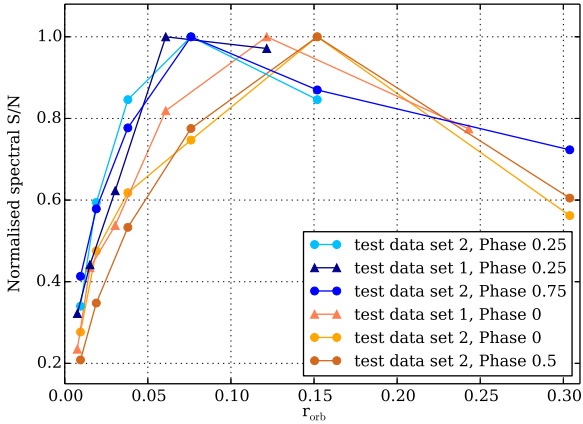}
\caption{Detected S/N versus $r_{\rm{orb}}$ for the two test data sets (test data set~1: triangle symbol, test data set~2: circle symbol) at the four selected orbital phases. Phases~0.25 and~0.75 represent orbital phases where $\dot{a}$ is largest. Phases~0 and~0.5 correspond to orbital phases where $\dot{a}\approx 0$.}
\label{fig:SNorbital}
\end{figure}

To determine the effectiveness of this constant acceleration approximation versus varying $r_{\rm{orb}}$, we employed two observations\footnote{We note that the two test data sets employed are taken on different dates with different instrumental set-ups, it is thus not appropriate to compare their detected S/N directly. However, one can attempt a qualitative comparison by normalising the highest S/N of any particular orbital phase to unity.} of the double pulsar system PSR~J0737$-$3039A as test data sets (refer to Appendix~\ref{app:rorb} for details of this study). Fig.~\ref{fig:SNorbital} is a plot of the S/N of four selected orbital phases from this study across varying $r_{\rm{orb}}$. From the plot it can be seen that a $r_{\rm{orb}}$ of roughly 0.1 can be adopted as a general rule-of-thumb in order to allow for an effective constant acceleration approximation, in agreement with the analytical sensitivity calculated in \citet{Ransom2003}. In addition, we point out a tendency for the orbital phases with a significant $\dot{a}$ to prefer shorter $r_{\rm{orb}}$ and vice versa for the orbital phases with an $\dot{a}\approx 0$. The introduction of eccentricity in the orbital motion will alter this picture. When the line of sight $\dot{a}$ is significantly non-zero, we expect smaller $r_{\rm{orb}}$ to perform better (i.e., the peaks of these curves shift towards the left). Otherwise, in the less accelerated part of the eccentric orbit, slightly larger $r_{\rm{orb}}$ can result in a higher S/N. For a quantification of the detectability of eccentric binary pulsars see e.g., \citet{Bagchi2013}.

\subsubsection{Acceleration ranges} \label{sec:AccRange}
\begin{figure*}
\centering
\setlength\fboxsep{-2pt}
\setlength\fboxrule{-2pt}
\includegraphics[width=14cm]{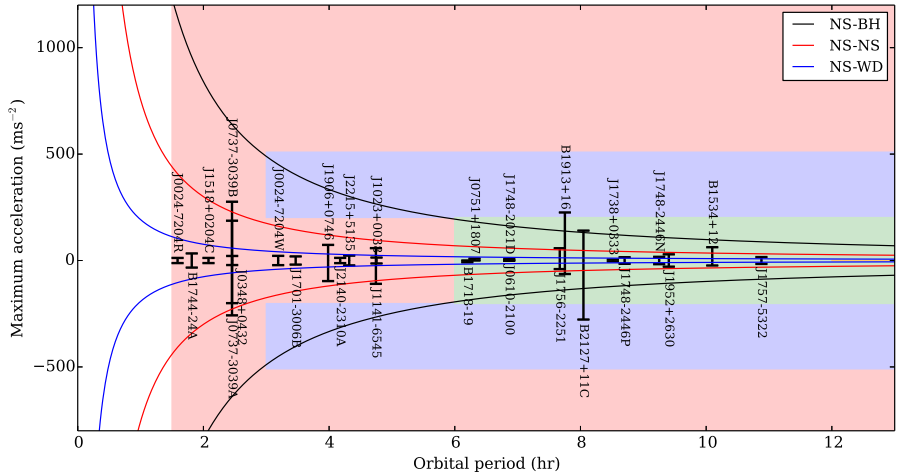}
\caption{Maximum orbital acceleration versus orbital period assuming circular orbits for binary systems of NS-WD (blue line), NS-NS (red line) and NS-BH with a 10\,$M_{\sun}$ BH (black line). The maximum orbital acceleration for all published relativistic binaries are also shown as a reference. The coloured regions correspond to the parameter spaces probed by different segments as explained in Section~\ref{sec:Segmentation}.} 
\label{fig:accrange}
\end{figure*} 

Assuming a circular orbit, we derive the theoretical maximum orbital acceleration ($a_{\rm{max}}$) for a given $P_{\rm{orb}}$ by applying Kepler's third law. We consider the upper limit case by setting the orbital inclination to be edge on (i.e., $i=90^{\circ}$):
\begin{equation}
\abs*{a_{\rm{max}}} = \left(\frac{2\pi}{P_{\rm{orb}}} \right)^{2} x c = \left(\frac{2\pi}{P_{\rm{orb}}} \right)^{4/3} (T_{\sun}f)^{1/3} c\,,
\label{eq:maxA}
\end{equation}
where $T_{\sun} = GM_{\sun}/c^{3} =4.925490947\,\upmu\rm{s}$, $x$ is the projected semi-major axis of the pulsar orbit and $c$ is the speed of light. The mass function $f$ is defined by
\begin{equation}
f = \frac{ (m_{\rm{c}} \sin{i})^{3} }{ ( m_{\rm{p}} + m_{\rm{c}} )^{2}}\,,
\label{eq:massfunction}
\end{equation}
where $\sin{i}=1$ and the pulsar mass, $m_{\rm{p}}$, is taken to be 1.4\,$M_{\sun}$. The companion mass, $m_{\rm{c}}$, then remains the only variable. Fig.~\ref{fig:accrange} shows the theoretical maximum orbital acceleration as a function of orbital period. We have plotted three scenarios corresponding to binary systems with a 0.2\,$M_{\sun}$ helium white dwarf companion (NS-WD), with a second 1.4\,$M_{\sun}$ neutron star companion (NS-NS) and with a hypothetical black hole companion of mass 10\,$M_{\sun}$ (NS-BH). We overplot all published relativistic pulsar binary systems with $P_{\rm{orb}}$ less than 12\,hr and significant orbital acceleration reaching above $\pm1$\,m\,s$^{-2}$ as a reference. 

Based on Fig.~\ref{fig:accrange}, sensible acceleration ranges ($\Delta{a}$) can be determined for any particular orbital period. We note that the effect of orbital eccentricity has not been taken into account. Highly-eccentric relativistic binary systems can overshoot these theoretical curves of maximum acceleration significantly, in particular during the orbital phase of periastron. Notable examples in Fig.~\ref{fig:accrange} are two NS-NS systems PSRs~B1913$+$16 and B2127$+$11C with orbital eccentricities of 0.62 and 0.68 respectively. To determine the theoretical maximum acceleration while including eccentric systems would dramatically increase the acceleration search parameter space, making the data processing unfeasible given the current computing resources available. Fortunately, a simple consideration of the Kepler's third law shows that, these eccentric systems spend only a relatively brief moment during the highly-accelerated orbital phase near periastron, whereas the majority of their orbital phases are confined within the less accelerated regime. Therefore, we justify that our acceleration ranges remain a reasonable compromise given the computing resources available.

\subsubsection{Partially-coherent segmentation} \label{sec:Segmentation}
\begin{figure*}
\centering
\setlength\fboxsep{0pt}
\setlength\fboxrule{0pt}
\fbox{\includegraphics[width=18cm]{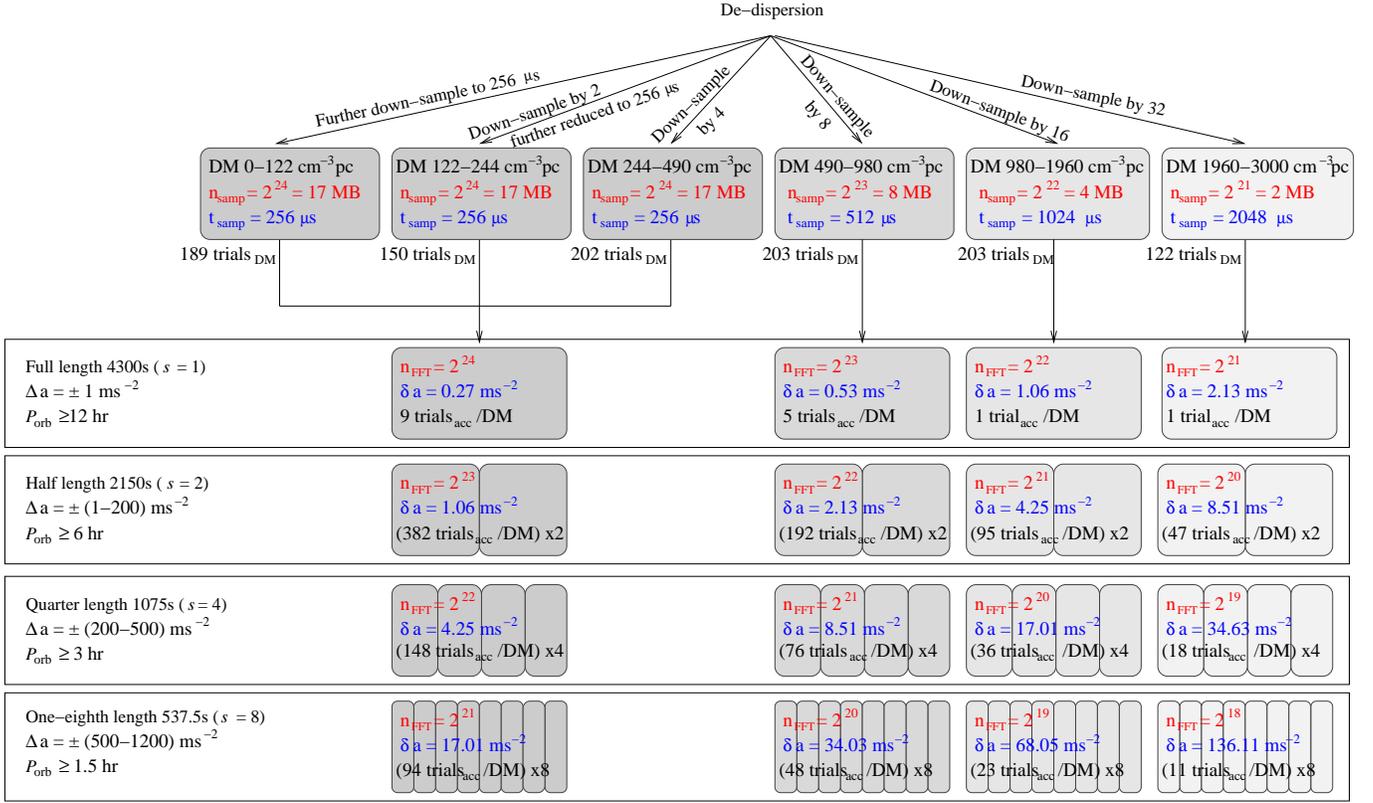}}
  \caption{A schematic diagram for the `partially-coherent segmented' pipeline adopted for this survey. As we progressively downsample the observation after $2\times$ the diagonal DM (DM = 122\,cm$^{-3}$\,pc) is reached, we end up with six groups of DM ranges with different $t_{\rm{samp}}$ (illustrated by the colour density in this schematic) each with different number of DM trials (trials$_{\rm{DM}}$). Each of the subsequent time series is then passed to the four lower panels (corresponding to the four configurations of $s=1,2,4,8$) to be Fourier transformed, where the number of samples used in each Fourier transform is marked as $n_{\rm{FFT}}$. We show also the acceleration range, $\Delta{a}$, the acceleration step size, $\delta{a}$, as well as the resultant number of acceleration trials per DM, trials$_{\rm{acc}}$/DM.} 
  \label{fig:flowchart}
\end{figure*}

As established in Section~\ref{sec:onetenth}, the constant acceleration approximation is most effective at $r_{\rm{orb}} \approx 0.1$. Hence in order to remain sensitive to a wide range of binary orbits it is strategic to use different integration lengths. Taking this as a rule-of-thumb, the 4300\,s full-length observation ($s=1$) of the HTRU Galactic plane survey would enable us to detect binary systems with $P_{\rm{orb}} \gtrsim 12$\,hr. Halving the observation into two equal segments ($s=2$) will correspond to binaries of $P_{\rm{orb}} \gtrsim 6\rm\,hr$, quartered-length observation ($s=4$) will correspond to $P_{\rm{orb}} \gtrsim 3\rm\,hr$ and segmenting our observation to one-eighth ($s=8$) will correspond to $P_{\rm{orb}} \gtrsim 1.5\rm\,hr$. We set aside binary systems with shorter $P_{\rm{orb}}$ for future re-processing. This is because given the long coherent integration length of this survey, such tight binaries should be most efficiently detected using the `phase modulation' technique described in \citet{Ransom2003}.

We search for binary systems using an acceleration range appropriate for the length of each segment, as discussed in Section~\ref{sec:AccRange}. The maximum acceleration attainable from a 10\,$M_{\sun}$ NS-BH system is of the order of 1200\,m\,s$^{-2}$ with a $P_{\rm{orb}}$ of 1.5\,hr, hence we have adopted this as the upper bound of our acceleration range for the shortest $s=8$ segments. The corresponding parameter space probed is shown as the pink region in Fig.~\ref{fig:accrange}. In order to maximise our detectability towards a NS-BH binary, we additionally search a $\Delta{a}$ between $\pm(200-500)$\,m\,s$^{-2}$ with the $s=4$ segments (providing sensitivity in the blue region), as well as a $\Delta{a}$ between $\pm(1-200)$\,m\,s$^{-2}$ with the $s=2$ segments (providing sensitivity in the green region). For the full length $s=1$ observation we search a $\Delta{a}$ of $\pm1$\,m\,s$^{-2}$, which should allow the detection of all mildly accelerated binary systems with large orbital periods of $P_{\rm{orb}} \ge 12$\,hr, as well as all isolated pulsar systems.

A schematic of the final pipeline is illustrated in Fig.~\ref{fig:flowchart}. We call it the `partially-coherent segmented acceleration search', as each segment is analysed coherently while across the segments the results are interpreted independently. We search the four configurations ($s=1,2,4,8$) in parallel, which is essentially equivalent to conducting a multiple pass survey. On one hand, an additional advantage of this partially-coherent scheme is that by independently analysing segments we are less susceptible to some epochs where the detection of the pulsar is more difficult, for instance at an orbital phase where $\dot{a}$ is significantly non-zero in a highly-eccentric orbit, an intermittent pulsar in switching off phase or a scintillation induced reduction in S/N. On the other hand, we note that this scheme relies on the fact that any binary pulsar to be found has to be detectable in at least one of the shortened segments. To be able to coherently combine the acceleration search results across segments would allow us to exploit the full sensitivity achievable with the deep integration of this survey, detecting even the weakest relativistic binary systems. This is the primary goal of our future data re-processing.

\section{Re-detections of known pulsars} \label{sec:known}
To verify that no previously known pulsar has been missed by the survey, we compute the expected S/N of every pulsar for any particular observation using the radiometer equation:
\begin{equation}
{\rm{S/N}}_{\rm{exp}} = \left(\frac { G \sqrt{  n_{\rm{p}} t_{\rm{int}} \Delta f }} { \beta T_{\rm{sys}}} \right)  S_{\rm{exp}} \left(\frac{W_{50}}{P-W_{50}}\right)^{-1/2}\,.
\label{eq:SNExp}
\end{equation}
The first fraction of Equation~(\ref{eq:SNExp}) contains parameters related to the observational set-up. The `degradation factor' $\beta$ is due to digitisation and is $\sim$1.16 for our case. The system temperature, $T_{\rm{sys}}$, includes contribution from both the sky temperature $(T_{\rm{sky}} \approx 7.6\,\rm{K})$ and the receiver temperature $(T_{\rm{rec}} \approx 23\,\rm{K})$, while $G$ is the antenna gain in K\,Jy$^{-1}$ and its value varies between 0.581, 0.690 and 0.735 depending on the receiver beam \citep[see Table~3 in][]{HTRU1}. The number of polarisations summed, $n_{\rm{p}}$, is always 2 in our case. The integration time, $t_{\rm{int}}$, is 4300\,s and $f$ is the effective bandwidth of the receiver which is 340\,MHz.

The remaining terms are pulsar dependent and we have taken the published data for these terms from \textsc{psrcat}. These parameters include the spin periods, $P$, the pulse widths at 50\,per cent, $W_{50}$, as well as the flux densities as observed at 1.4\,GHz. Note that we take into account the relationship between the reduction in expected flux density, $S_{\rm{exp}}$, and the catalogue flux density, $S_{1400}$, if the pulsar is offset from the beam centre. This is given by
\begin{equation}
S_{\rm{exp}} = S_{1400} \exp \left( -\frac{\theta^2}{2\sigma^2} \right)\,,
\label{eq:FluxExp}
\end{equation}
where $\theta$ is the radial distance between the published pulsar position and the centre of the relevant beam. Assuming a Gaussian drop-off of beam sensitivity with a FWHM of 14$'$.4, we define $\sigma$ to be:
\begin{equation}
\sigma = \frac{\rm{FWHM}}{2 \sqrt{2\ln 2} } \approx 0.1\,\degr \,.
\label{eq:FWMH}
\end{equation}

\begin{figure}
\centering
\includegraphics[width=3in]{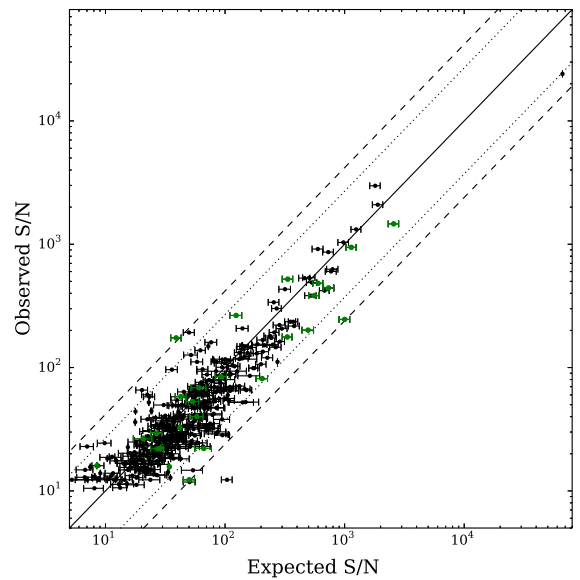}
\caption{A comparison of the observed S/Ns of known pulsar re-detections with their predicted values. The solid line shows the expected 1:1 correlation. The dotted and the dashed lines mark the region containing 95 and 99 per cent of the data points. Pulsars with DM$\le$100\,cm$^{-3}$\,pc are potentially affected by scintillation and are coloured in green. The error bars reflect the published uncertainties in the pulsar flux densities from \textsc{psrcat}.}
\label{fig:SNcomp}
\end{figure}

The minimum detectable S/N is based on the false alarm statistics \citep[see e.g.,][]{Handbook2004} and for this survey it is taken to be 9. Within the \PROCESSED{} of processed HTRU Galactic plane observations, we record \REDETECTIONu{} independent known pulsar re-detections from \REDETECTIONp{} pulsars (refer to a re-detection table available as online material, a sample of which can be see in Table~\ref{tab:app} in the Appendix). 

Fig.~\ref{fig:SNcomp} plots the observed S/N$_{\rm{obs}}$ versus the expected (S/N$_{\rm{exp}}$) calculated from Equation~(\ref{eq:SNExp}) for the known pulsar re-detections. Out of the \REDETECTIONu{} re-detections, \SIDELOBE{} lie outside the $14'.4$ FWHM of the receiver beam. Note that the sensitivity of the telescope outside the FWHM of the primary beam is complicated by its sidelobe pattern and is not well quantified. Hence these \SIDELOBE{} re-detections are disregarded for the purpose of S/N comparison. Another five detections come from known pulsars with no published catalogue flux density. They are therefore not included in Fig.~\ref{fig:SNcomp}. The black points in Fig.~\ref{fig:SNcomp} denote the remaining \INBEAM{} known pulsar re-detections within the FWHM. It can be seen that the correlation between S/N$_{\rm{exp}}$ and S/N$_{\rm{obs}}$ is very good. Most of the data points lie close to the line of 1:1, implying the wider $\Delta{f}$ and the longer $t_{\rm{int}}$ of the HTRU survey has provided the increased sensitivity expected. 

\citet{HTRU1} have pointed out that data points below the 1:1 correlation could be influenced by the bias resulting from the tendency to publish the discovery observation which is more likely to have the highest S/N due to scintillation, leading to the higher catalogue flux densities. Despite this, a few re-detections are significantly worse than expected and lie below the 99~per cent lower boundary. They are likely to be due to scintillation and/or contamination by RFI. In the case of the very bright pulsars on the top right-hand corner of Fig.~\ref{fig:SNcomp}, the loss in survey sensitivity might be a consequence of the 2-bit digitisation. This, however, is not a problem for our purpose here.

Out of all the expected known pulsar re-detections, \MISSED{} have been missed in the data processing (refer to Table~\ref{tab:missedPSR} in the Appendix for potential reasons of these non-detections). Folding the observation directly with the catalogue pulsar ephemerides resulted in the recovery of seven pulsars (listed as S/N$_{\rm{eph}}$ in Table~\ref{tab:missedPSR}). Among the missed re-detections, one is attributed to the magnetar PSR~J1622$-$4950. This magnetar is being followed up by an on-going HTRU timing programme. However, its radio emission appears to have ceased since October~2010, which is consistent with the non-detection of the HTRU Galactic plane survey observation taken in December~2010. The radio emission of magnetars are known to fade with time, hence we have likely observed the end of the radio emission of PSR~J1622$-$4950. Excluding several other similar cases (refer to Appendix~\ref{app:missed} for the reasoning) where phenomena intrinsic to the pulsar have prevented the re-detections, we conclude that only $\sim1$~per cent of the published pulsars expected have been missed and the large number of re-detections indicate that the HTRU Galactic plane survey is performing as expected with little loss of sensitivity. We also remark that none of the missed pulsars belongs to binary systems (refer to Table~\ref{tab:knownbinary} in the Appendix).

\section{Newly-discovered pulsars} \label{sec:new}
\begin{table}
\setlength{\tabcolsep}{0.11cm}
 \caption{Table listing the S/N, flux density ($S_{1400}$), derived luminosity ($L_{1400}$) and pulse widths ($W_{50}, W_{10}$) of the \NEWPSR{} newly-discovered pulsars.}  
  \begin{tabular}{llllllll}
  \hline
  PSR~Jname  & S/N & S/N & S/N & $S_{1400}$ & $W_{50}$ & $W_{10}$  & $L_{1400}$\\
             & $_{\rm{HTRU}}$ & $_{\rm{medlat}}$  & $_{\rm{PMPS}}$ & (mJy)      & (ms)     & (ms)      & (mJy\,kpc$^{2}$) \\
\hline
J1002$-$5919 & 18.2 & 6.8 & 7.7  & 0.17 & 61.3 & 69.1 & 12.6 \\
J1101$-$6424 & 19.7 & 7.4 & $-$  & 0.27 & 0.54 & 1.39 & 5.4 \\
J1151$-$6108 & 10.0 & $-$ & $-$  & 0.06 & 5.65 & 9.89 & 1.1 \\
J1227$-$63   & 19.4 & $-$ & $-$  & $-$  & 15.0 & 34.4 & $-$ \\
J1244$-$6359 & 12.5 & $-$ & $-$  & 0.15 & 9.8  & 22.9 & 4.7 \\
J1248$-$6444 & 13.4 & $-$ & $<6.4$& 0.15 & 54.3 & 78.5& 3.3 \\
%
J1255$-$62   & 14.2 & $-$ & $-$ & 0.13  & 34.8 & 65.0 &  21.8 \\
J1349$-$63   & 17.4 & $-$ & 9.7 & $-$   & 23.0 & 49.7 &  $-$\\
J1525$-$5523 & 12.6 & 10.3& 10.5& 0.21 & 31.0 & 40.9 &  1.1 \\
J1528$-$5547 & 14.4 & $-$ & $-$  & 0.07 & 52.9 & 110  &  1.2 \\
J1532$-$56   & 12.0 & $-$ & 9.7  & 0.10 & 27.0 & 54.0 &  1.8\\
J1538$-$5621 & 22.7 & 9.2 & 15.7 & 0.14 & 34.1 & 60.6 &  1.2\\
%
J1612$-$49   & 12.6 & $-$ & $-$ & 0.16  & 28.7 & 73.1 &  10.3 \\
J1612$-$55   & 17.7 & $-$ & $-$ & 0.11  & 37.6 & 56.5 &  3.6  \\
J1616$-$5017 & 21.5 & $-$ & $-$ & 0.17  & 9.76 & 20.5 &  3.8 \\
J1622$-$4845 & 12.1 & $-$ & $-$ & 0.15  & 17.5 & 42.3 &  3.1 \\
J1627$-$49   & 14.1 & $-$ & 8.6 & 0.13  & 69.3 & 97.1 &  6.0 \\
J1627$-$51   & 21.6 & $-$ & 13.3& 0.08  & 9.8  & 16.3 &  0.98 \\   
%
J1634$-$49   & 14.3 & $-$ & 8.5 & 0.13  & 43.5 & 62.5 &  10.5 \\
J1638$-$44   & 15.5 & $-$ & $-$ & 0.15  & 31.6 & 50.6 &  6.8 \\
J1649$-$3935 & 17.2 & 6.8 & 9.6 & 0.05  & 22.8 & 45.7 &  1.6 \\
J1658$-$47   & 11.5 & 6.9 & 7.7 & 0.18  & 27.4 & 43.8 &  23.0 \\
J1708$-$3641 & 13.5 & $-$ & $<6.5$& 0.12& 21.8 & 78.3 &  2.7 \\
J1710$-$37   & 13.3 & $-$ & $<7.7$& 0.10& 61.4 & 105  &  2.2\\
J1718$-$41   & 12.4 & $-$ & $-$  & 0.07 & 28.4 & 69.1 & 2.0 \\
J1720$-$36   & 14.5 & $-$ & $-$  & 0.08 & 8.41 & 10.8 & 1.4 \\
J1723$-$38   & 14.3 & $-$ & 10.6 & 0.08 & 13.7 & 35.0 & 2.8 \\
J1730$-$34   & 15.4 & $-$ & 7.1  & 0.03 & 25.9 & 51.7 & 1.5 \\
J1731$-$3322 & 10.5 & $-$ & $<6$ & 0.10  & 33.5 & 71.4 & 12.5 \\
J1732$-$35   & 11.1 & $-$ & $-$  & 0.13  & 6.6  & 14.1 & 3.3 \\
%
J1734$-$3058 & 21.0 & $-$ & 9.3 & 0.11  & 10.7 & 22.6 & 1.6\\
J1738$-$2736 & 26.0 & $-$ & 9.5 & 0.17  & 13.7 & 23.7 & 4.4 \\
J1741$-$34   & 12.0 & $-$ & $-$ & 0.20  & 19.1 & 43.4 & 3.4 \\
J1743$-$35   & 14.4 & $-$ & 9.7 & 0.05  & 16.9 & 33.8 & 0.5 \\
J1746$-$27   & 8.7  & $-$ & $-$ & 0.15  & 29.9 & 42.6 & 5.6 \\
J1748$-$30   & 12.3 & 6.0 & 8.9 & 0.16  & 25.5 & 79.4 & 10.5 \\
%
J1750$-$28   & 11.6 & $-$ & $-$ & 0.09  & 16.8 & 35.1 &  2.3 \\
J1755$-$25   & 15.1 & $-$ & 9.0 & 0.14  & 11.7 & 25.7 &  14.9 \\
J1755$-$26   & 15.6 & 6.9 & $-$ & 0.14  & 16.3 & 25.7 &  3.8\\
J1756$-$25   & 20.5 & $-$ & $-$ & 0.20  & 25.4 & 38.0 &  18.4 \\
J1757$-$27   & 14.6 & $-$ & $-$ & 0.07  & 0.49 & 0.88 &  2.0\\
J1759$-$24   & 28.5 & 7.6 & $-$ & 0.50  & 72.3 & 159  &  59.4 \\
%
J1811$-$1717 & 14.2 & 6.4 &$<5.8$& 0.20  & 38.8 & 59.0 & 9.0\\
J1819$-$17   & 11.2 & $-$ & $-$  & 0.14  & 87.1 & 157  & 0.5 \\
J1824$-$1350 & 14.6 & $-$ & $-$  & 0.08  & 19.4 & 58.2 & 6.5 \\
J1825$-$1108 & 17.5 & $-$ & 7.5  & 0.13  & 42.1 & 68.8 & 0.9 \\
J1829$-$1011 & 19.8 & $-$ & 9.3  & 0.25  & 98.8 & 125  & 11.2 \\
J1830$-$10   & 9.9  & $-$ &$<7.3$& 0.10  & 5.36 & 12.7 & 1.3 \\
%
J1833$-$0209 & 11.9 & $-$ &$<6.8$& 0.06  & 17.4 & 30.7 &  3.1 \\
J1834$-$09   & 18.1 & $-$ & $-$  & 0.17  & 64.5 & 152  &  7.0 \\
J1835$-$0924b& 13.0 & 8.9 & $-$  & $-$   & 27.1 & 45.3 &  $-$ \\
J1835$-$0928 & 10.5 & $-$ & $-$  & $-$   & 39.5 & 64.2 &  $-$ \\
J1836$-$11   & 10.8 & $-$ & $-$  & 0.12  & 25.7 & 55.1 &  3.2 \\
J1838$-$0107 & 27.1 & 10.5& 9.7  & 0.05  & 34.4 & 45.0 &  1.9 \\
J1839$-$0223 & 11.2 & $-$ & 8.3  & 0.11  & 45.3 & 67.9 &  4.5 \\
J1839$-$0332 & 20.2 & 8.6 & 12.7 & $-$   & 31.9 & 63.7 &  $-$ \\
J1842$-$0800 & 9.8  & $-$ & 9.9  & 0.10  & 24.9 & 52.3 &  1.7 \\
J1843$-$0510 & 11.8 & $-$ & 7.5  & 0.07  & 24.0 & 36.0 &  1.9 \\
J1844$-$0302 & 13.6 & $-$ & $-$  & 0.12  & 19.0 & 33.3 &  6.4 \\
J1847$-$0427 & 11.4 & $-$ & $-$  & $-$   & 30.9 & 48.4 &  $-$ \\
  \hline \label{tab:newSN}
 \end{tabular}
\end{table}

A total of \NEWPSR{} pulsars have been discovered in the {\PROCESSED}{} processed data of the Galactic Plane survey. Listed in Table~\ref{tab:newSN} are their folded signal-to-noise ratios (S/N$_{\rm{HTRU}}$) at discovery. The previous PMPS as well as the medium-latitude sub-survey of the HTRU have a complete overlap in the region of sky with the HTRU Galactic plane survey. We have inspected the PMPS and medium-latitude archival data to determine if any of the newly-discovered pulsars were detectable and listed the respective folded S/N$_{\rm{PMPS}}$ and S/N$_{\rm{medlat}}$ in columns~3 and~4 of Table~\ref{tab:newSN}. Eighteen and five pulsars have respective S/N$_{\rm{PMPS}}$ and S/N$_{\rm{medlat}}$ greater than the theoretical detection threshold of S/N$_{\rm{min,PMPS}}=$S/N$_{\rm{min,medlat}}=8$, which means they could have been discovered in these archival observations. They might have been missed due to the large number of candidates produced in the PMPS and medium-latitude processing. Improvements in the RFI mitigation scheme of the current HTRU processing as mentioned in Section~\ref{sec:RFI} have likely helped avoiding such candidate confusion, enabling the detection of these pulsars. A further twelve and eight pulsars have weak detections from these archival observations, with respective S/N$_{\rm{PMPS}}$ and S/N$_{\rm{medlat}}$ less than 8. In the cases when only tentative detections are suggestive we have listed an upper limit for their S/N$_{\rm{PMPS}}$ or S/N$_{\rm{medlat}}$ in Table~\ref{tab:newSN}. These pulsars might have been missed in the processing of these archival surveys as their low S/Ns might have prevented them from being selected for visual inspection. The remaining 25 pulsars are detectable from neither the PMPS nor the medium-latitude archival data.

\begin{table}
\setlength{\tabcolsep}{0.13cm}
 \caption{Specifications of the observing system employed for the timing observations in this work. The antenna gain is represented by $G$ and $T_{\rm{sys}}$ is the receiver system temperature. The central frequency in MHz is represented by $f_{\rm{c}}$ and $B$ is the bandwidth in MHz.}
\begin{tabular}{lp{1.1cm}p{0.6cm}lp{0.8cm}p{0.9cm}}
  \hline
  Receiver  & $G$ $\rm(K\rm\,Jy^{-1})$ & $T_{\rm{sys}}$ (K) & Backend       & $f_{\rm{c}}$ (MHz) & $B$ (MHz)  \\
  \hline
  Multibeam & 0.74 & 23 & Parkes DFBs   & 1369               & 256        \\
            &       &    & Parkes BPSR   & 1352               & 340        \\
            &       &    & Parkes APSR   & 1369               & 256       \\
  Single-pixel & 1.00   & 28   & Lovell DFB  & 1532                & 384        \\
            &    &    & Jodrell ROACH & 1532               & 400       \\
  \hline \label{tab:specs}
 \end{tabular}
\end{table}

Full flux density calibration is implemented for \CALIB{} pulsars with observations taken with the Parkes Digital Filter bank systems (DFB). We calibrate each observation by using an averaged observation of Hydra A, and we account for the differential gain and phase between the feed with an observation of the noise diode coupled to the the feeds. It is important that this calibration is taken adjacent to the targeted pulsar observations. Column~5 of Table~\ref{tab:newSN} reports the mean flux densities averaging over all available timing observations for each pulsars at 1.4\,GHz, $S_{1400}$.  In turn, we infer the luminosity at 1.4\,GHz, $L_{1400} = S_{1400} \times d^{2}$, where $d$ is the distance of the pulsar in kpc (see column 9 of Table~\ref{tab:timing1}) in accordance to the NE2001 electron density model \citep{NE2001model}.

Also listed in columns~6 and~7 of Table~\ref{tab:newSN} are the pulse widths measured at 50 ($W_{50}$) and 10 ($W_{10}$) per cent of the highest peak. The pulse profile of each pulsar at 1.4\,GHz is shown in Fig.~\ref{fig:profilenew}. For the pulsars with coherent timing solutions, we summed all observations to form high S/N mean profiles. Otherwise, we plot the profiles from the single observation at discovery. The peak of the profiles have been normalised to unity and their peaks placed at phase 0.2. Most of the long-period normal pulsars have typical pulse profiles \citep{Lyne2005}, with single-peaked pulses and $\delta < 10$~per cent. One of the \NEWPSR{} newly-discovered pulsars, PSR~J1847$-$0427, shows a broad pulse profile with a hint of an interpulse and is further discussed in Section~\ref{sec:J1847}. These results are entirely consistent with the findings of \citet{Weltevrede2008}, that $\sim$2~per cent of the known pulsar population is observed with an interpulse. A few of the pulsars with high DM display the classical exponential tail of scattering caused by propagation of the radio signal through the interstellar medium. 

When a pulsar is first discovered, our knowledge of its sky position, rotation period and DM are only approximate. Follow up timing observations of at least 1\,yr are necessary to precisely determine its rotational, astrometric and, if any, orbital parameters. We sum each timing observation over both frequency and time to produce an integrated pulse profile. The \textsc{psrchive} data analysis package \citep{Hotan2004} is used to convolve a noise-free analytic reference template with each individual profile to produce a time of arrival \citep[TOA;][]{Taylor1992}. The \textsc{tempo2} software package \citep{Hobbs2006} is then employed to fit a timing model to all TOAs of the pulsar. Towards the end of the timing analysis procedure when the respective reduced $\rm\chi^2$ comes close to one, we can then assume a reliable fit is achieved which is only influenced by the presence of radiometer noise in the template. As a last step, we compensate for these systematic effects by calculating dataset-specific calibration coefficients (also known as `EFAC' in {\textsc{tempo2}). These coefficients are applied to scale the TOA uncertainties such that each final respective reduced $\rm\chi^2$ is unity.

All observations presented here have been taken at 1.4\,GHz at Parkes using backends including the DFBs with incoherent dedispersion and the ATNF Parkes Swinburne Recorder\footnote{http://astronomy.swin.edu.au/pulsar/?topic=apsr} (APSR) with coherent dedispersion \citep{Hankins1975}. Pulsars with declination North of $-35\degr$\footnote{In practice, for the declination between $-30\degr$ and $-35\degr$, only bright pulsars are followed-up at the Lovell Telescope. This is because of the short visible hours, as well as the challenges associated with such low elevation observations, namely the need for low wind conditions and the potential high spillover.} were timed at the Jodrell Bank Observatory with the Lovell 76-m telescope, using a DFB backend and a ROACH backend. The latter is based on the ROACH FPGA processing board\footnote{https://casper.berkeley.edu/wiki/ROACH} and coherently dedisperses the data. Refer to Table~\ref{tab:specs} for the specifications of all observing systems employed. Timing observations at Jodrell Bank were performed approximately once every three weeks, whereas Parkes observations are more irregular with gaps ranging from days to months depending on telescope availability. At both telescopes, integration times of at least 20\,min are typically required, with longer observations for weaker pulsars to achieve adequate S/N of at least 10. 

Table~\ref{tab:newPSR} presents the \NEWSOLUTIONiso{} newly-discovered isolated pulsars with coherent timing solutions, whereas the \NEWSOLUTIONbin{} newly-discovered binary systems are listed in Table~\ref{tab:solutionBinary}. The Damour-Deruelle (DD) timing model \citep{DD1986} in \textsc{tempo2} is a theory-independent description for eccentric binary orbits. However, for binaries with small eccentricities the location of periastron is not well-defined and using the DD timing model results in a high covariance between the longitude of periastron $(\omega)$ and the epoch of periastron $\rm(T_{0})$. For thse pulsars, we use the ELL1 timing model \citep{Lange2001} alternatively. The ELL1 timing model avoids the covariance by using the Laplace-Lagrange parameters ($\epsilon_{1}=e\sin\omega$ and $\epsilon_{2}=e\cos\omega$) and the time of ascending node passage $(\rm{T_{asc}})$ instead of $\rm{T_{0}}$ as in the DD timing model. A remaining \NEWNOSOLUTION{} of the most recent discoveries have not yet been allocated sufficient follow-up telescope time, but will be monitored in the coming months. For these pulsars we have reported their discovery parameters in Table~\ref{tab:newPSR2}. 

No associated gamma-ray pulsations have been identified for any of the newly-discovered pulsars presented in this paper. Currently, the short time span of the radio ephemerides are inadequate to phase-fold the now more than five years data from the Large Area Telescope (LAT) on the {\em Fermi Gamma-Ray Space Telescope}. Continuous follow-up timing of these pulsars and any future discoveries from the HTRU Galactic plane survey would provide up-to-date radio ephemerides critical for recovering any associated gamma-ray pulsations. We note that for the \PPDOT{} pulsars currently with $P$ and $\dot{P}$ measurements, none of them has high $\log\sqrt{\dot{E}}/d^{2}$ \citep[see ][for details]{Fermi2PC}, hence unlikely to be detected by {\em Fermi} in the future.

\begin{figure*}
\centering
\setlength\fboxsep{0pt}
\setlength\fboxrule{0pt}
\fbox{\includegraphics[width=16cm]{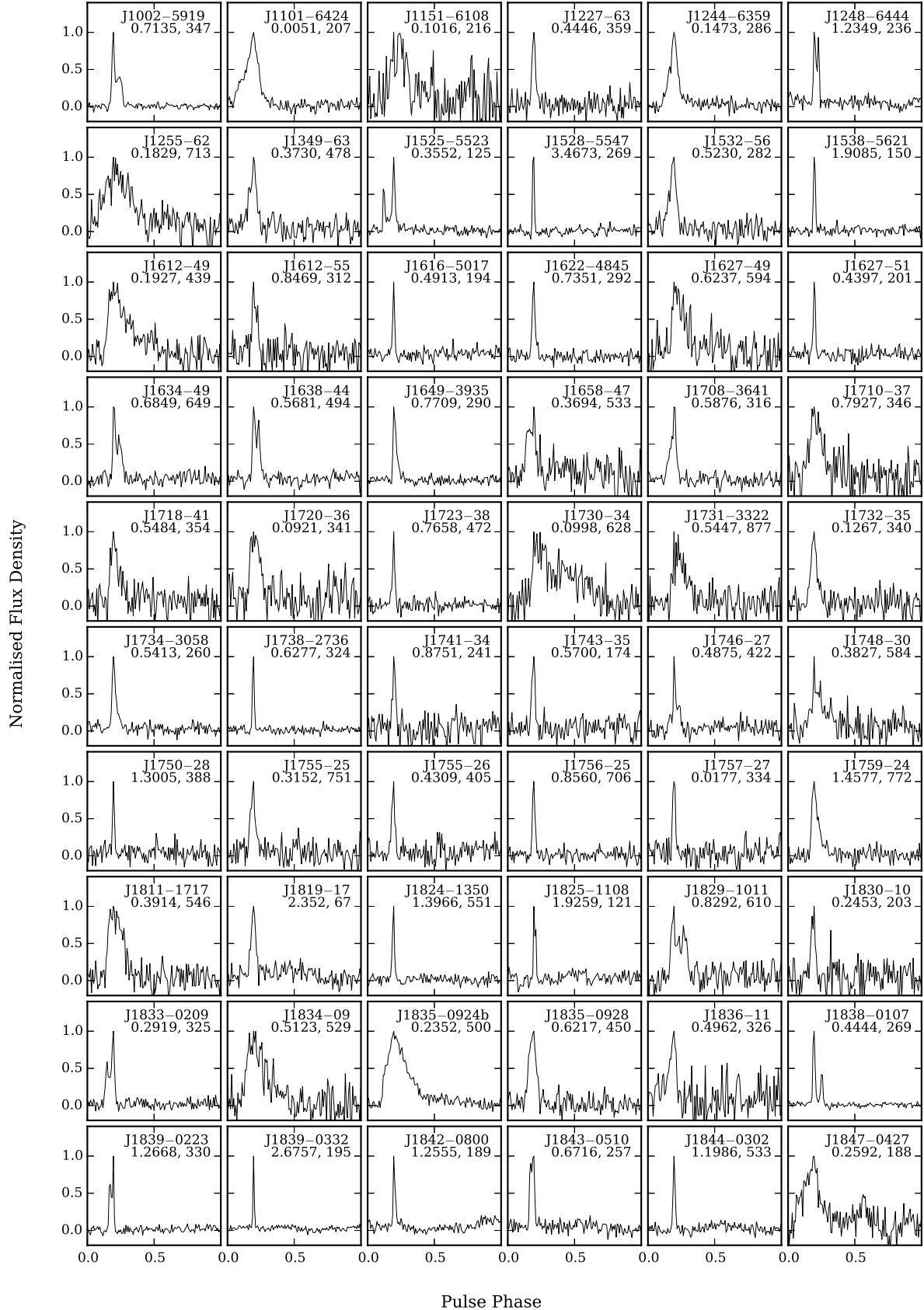}}
  \caption{Average pulse profile of the {\NEWPSR}{} newly-discovered pulsars, plotted by placing the peak at phase 0.2 and scrunching to 128 phase bins. Also marked on each profile is the respective pulsar names, the spin periods in s followed by the DMs in cm$^{-3}$\,pc.}
  \label{fig:profilenew}
\end{figure*}

\begin{landscape}
\begin{table}
\centering
\setlength{\tabcolsep}{0.07cm}
\caption{\textsc{tempo2} best-fitting parameters of \NEWSOLUTIONiso{} newly-discovered isolated pulsars from the HTRU Galactic plane survey. We list the equatorial (R.A. and Dec.) and Galactic ($l$ and $b$) position, the spin period and the DM of these pulsars. Values in parentheses are the nominal 1$\sigma$ uncertainties in the last digits. Pulsars for which no full timing solution is available have been assigned a temporary name containing only two digits of declination. Pulsar distances are derived according to \protect \citet{NE2001model}. We include fitting related parameters such as the data span, the reference epoch, the number of TOAs employed (N$_{\rm{TOA}}$), the RMS of the \textsc{tempo2} fit and the reduced $\chi^{2}$ ($\chi^{2}_{\rm{red}}$). The characteristic age ($\tau_{\rm{c}}$), the surface magnetic field ($B_{\rm{surf}}$) and the spin-down energy ($\dot{E}$) are derived using equations which can be found in \protect\citet{Handbook2004}. }  \label{tab:timing1}
\resizebox{\columnwidth}{!}{%
 \begin{tabular}{lllR{0.8cm}R{0.8cm}lllllllllllp{1cm}}
  \hline
 PSR~name & R.A. (J2000)  & Dec. (J2000)   & $l$ & $b$ & $P$ & $\dot{P}$ &
 DM & Dist & Data span  & Epoch & $N$ & RMS & $\chi^{2}_{\rm{red}}$ $^{\ddagger}$ & $\tau_{\rm{c}}$ & $B_{\rm{surf}}$ & $\dot{E}$ \\
  & $(^{\rm{h m s}})$ & $(^\circ\,'\,'')$ & $(^{\circ})$ & $(^{\circ})$ & (ms) & $(10^{-18})$ &
(cm$^{-3}$pc) & (kpc) & (MJD) &(MJD)  & $_{\rm{TOA}}$ & ($\upmu$s) & & (Myr) & $(10^{10}$G) & $(10^{30}\rm{erg}\,\rm{s}^{-1})$ \\
  \hline
J1002$-$5919 & 10:02:20.50(5)  & $-$59:19:37.2(6) & 282.83 & $-$3.22 &  713.4892535(6)    & 104(10)  & 347(2) & 8.7 & 56499$-$56930 &  56677 & 28 & 1262 & 1.3 & 110 & 28 & 14 \\ 
J1151$-$6108 & 11:51:56.86(2) & $-$61:08:17.6(3) & 295.81 & 0.91    & 101.633196293(2) & 10278.4(10) & 216(1) & 4.2 & 56499$-$56930 & 56714 & 23 & 696 & 1.5 & 0.16 & 100 & 390000\\
J1248$-$6444 & 12:48:32.87(2) & $-$64:44:00.0(8) & 302.62 & $-$1.86 & 1234.89349389(3)  & 1932(16) & 236.4(11) & 4.8 & 56498$-$56931 & 56715 & 14 & 567 & 0.6 & 10 & 160 & 41 \\
J1525$-$5523 & 15:25:36.064(7) & $-$55:23:27.2(2) & 323.66 & 1.15 & 355.1560369998(16) & 8.96(7) & 124.7(3) & 2.3 & 54792$-$56571 & 55682 & 29 & 396 & 0.6 & 630 & 5.7 & 7.9 \\
J1528$-$5547 & 15:28:39.18(12) & $-$55:47:23(2) & 23.80 & 0.58 &  3467.3015170(17) & 7750(20) & 269 & 4.2 & 56127$-$56967 & 55839 & 27 & 7798 & 8.2 & 7.1 & 525 & 7.4 \\ 
J1538$-$5621 & 15:38:43.29(2) & $-$56:21:55.5(4) & 324.61 & $-$0.70 & 1908.49449143(2) & 7311.8(7) & 150(18) & 3.0 & 54939$-$56886 & 55912 & 14 & 926 & 0.9 & 4.1 & 380 & 42  \\
J1616$-$5017 & 16:16:29.89(4) & $-$50:17:14.9(2) & 332.83 & 0.29   & 491.384121830(13) & 46592(3)  & 194(4) & 4.7 & 56469$-$56933 & 56701 & 14 & 597 & 3.4 & 0.17 & 480 & 16000 \\
J1622$-$4845 & 16:22:05.45(10) & $-$48:46:29(4) & 334.53 & 0.73  & 735.09164794(8)   & 110(20) & 292(7) & 4.6 & 56499$-$56968 & 56715 & 12 & 1056 & 1.3 & 98 & 30 & 12 \\ 
J1649$-$3935 & 16:49:06.68(3) & $-$39:35:44.2(14) & 344.58 & 3.34    & 770.909760511(14)  & 39.1(4)  & 290(7) & 5.4 & 55085$-$56886 & 55985 & 15 & 1195 & 1.5 & 310 & 18 & 3.4 \\
J1708$-$3641 &  17:08:35.91(3) & $-$36:41:21.5(14) & 349.23 & 2.13 & 587.566793839(14) & 134.9(5) & 316(3) & 4.8 & 55208$-$56886 & 56047 & 19 & 1581 & 0.6 & 69 & 29 & 26 \\
J1731$-$3322 & 17:31:14.31(10) & $-$33:22:45(2) & 354.60 & 0.22 & 544.67054719(4) & 28093(18) & 877.4 & 11.0 & 56500$-$56932 & 56716 & 26 & 2931 & 1.5 & 0.3 & 400 & 6900 \\ 
J1734$-$3058  & 17:34:50.839(7) & $-$30:58:41.4(8) & 357.03 & 0.89 & 541.285680730(5) & 16.2(5)  & 260.0(14) & 3.8  & 55976$-$56784 & 56400 & 128 & 1750 & 2.8 & 530 & 9.5 & 4.0 \\
J1738$-$2736 & 17:38:14.628(5) & $-$27:36:25.8(9) & 0.27 & 2.08 & 627.715518484(19) &  4624.3(7)  & 323.6(7) & 5.1  & 56322$-$56884 & 56315 & 69 & 760 & 3.6 & 2.2 & 170 & 740\\
J1750$-$28 & 17:50:04.0(2) & $-$28:45(7) & 0.66   & $-$0.74 & 1300.5131462(7) & 5780(190)    & 388(12) & 5.1  & 56681$-$56884 & 56783 & 22 & 1567 & 1.7 & 3.6 & 280 & 100 \\
J1755$-$26 & 17:55:16.29(4) & $-$26:00(1) & 3.62   & $-$0.33 & 430.87215605(14) & 12280(50) & 405(4) & 5.2 & 56684$-$56940 & 56827 & 30 & 1207 & 2.4 & 0.55 & 230 & 6100 \\
J1757$-$27 & 17:57:54.795(2) & $-$27:45(7) & 2.40 & $-$1.72 & 17.68721477177(11) & 0.21(2) & 334 & 5.3 & 56675$-$56881 & 56778 & 50 & 24 & 2.0 & 1300 & 0.20 & 1600  \\
J1811$-$1717 & 18:11:26.53(19) & $-$17:17:47(14) & 13.04 & 0.69 & 391.38510099(9) & 120(80)  & 545.5 & 6.7 & 56640$-$56884 & 56762 & 39 & 2461 & 1.3 & 49 & 22 & 83 \\
J1824$-$1350 & 18:24:50.181(17) & $-$13:50:21.0(16) & 17.61 & $-$0.52 & 1396.59854604(4) & 607(7) & 551(7) & 6.5 & 56341$-$56884 & 56612 & 75 & 2957 & 1.9 & 36 & 93 & 8.8 \\
J1825$-$1108 & 18:25:18.3(2) & $-$11:08:54(12) & 20.05  & 0.64 & 1925.8717727(8) & 1900(600) & 121(19)  & 2.7 & 56682$-$56884 & 56783 & 29 & 1003 & 3.5 & 16 & 200 & 11\\
J1829$-$1011 & 18:29:06.06(4) & $-$10:11:25(4) & 21.33 & 0.26 & 829.16602078(16) & $-$5(39) & 610 & 6.7 & 56626$-$56980 & 56755 & 44 & 2962 & 0.7 & $>170^{\dagger}$  & $<25^{\dagger}$ & $<5.2^{\dagger}$ \\
J1833$-$0209 & 18:33:05.411(8) & $-$02:09:16.4(3) & 28.92 & 3.09 & 291.930632181(4) & 2754.8(7)  & 325.4 & 7.2 & 56341$-$56886  & 56614 & 97 & 1535 & 1.8 & 1.7 & 91  & 4400  \\
J1835$-$0924b$^{*}$ & 18:35:21.822(16) & $-$09:24:15.9(10) & 22.74 & $-$0.74 & 235.248924972(4) & 11.84(3)  & 500 & 6.3 & 55905$-$56885 & 56394 & 142 & 4576 & 2.7 & 315 & 5.3 & 36\\
J1835$-$0928 & 18:35:22.22(3) & $-$09:28:02(2) & 22.68 & $-$0.77 & 621.73399423(3) & 969(2) & 450 & 5.8 & 55906$-$56885 & 56395 & 98 & 6772 & 5.9 & 10 & 79 & 160  \\
J1838$-$0107 & 18:38:39.423(2) & $-$01:07:48.64(10) & 30.48 & 2.33 & 444.4257246169(17) & 5.46(15) & 268.9 & 6.1 & 56094$-$56884 & 56489 & 131 & 602 & 2.2 & 1300 & 5.0 & 2.5 \\
J1839$-$0223 &  18:39:58.032(3) & $-$02:23:09.5(16) & 29.50 & 1.46 &  1266.79012424(5) & 4763(6)  & 330 & 6.4 & 56094$-$56881 & 56488 & 117 & 9103 & 6.3 & 4.2 & 250 & 92 \\
J1839$-$0332 &  18:39:56.583(12) & $-$03:32:58.6(6) & 28.46 & 0.93&  2675.68226451(6) & 4760(6)  & 195.1 & 4.8 & 56098$-$56885 & 56491 & 101 & 2820 & 1.7 & 8.9 & 360 & 9.8\\
J1842$-$0800 & 18:42:54.943(13) & $-$08:00:53.5(7) & 24.83 & $-$1.77 &  1255.46857429(3) & 194(5)  & 188.6 & 4.1 & 56341$-$56880 & 56611 & 57 &  1949 & 0.9 & 100 & 50 & 3.8 \\
J1843$-$0510 & 18:43:09.732(12) & $-$05:10:04.5(5) & 27.39 & $-$0.52 & 671.613828819(13) & 3892.3(12)  & 257 & 5.2 & 56144$-$56885 & 56514 & 98 & 2572 & 2.0 & 2.74 & 160 & 510 \\
J1844$-$0302 & 18:44:06.918(12) & $-$03:02:11.2(5) & 29.40 & 0.24 & 1198.63027492(2) & 7809(5)  & 533 & 7.3 & 56341$-$56885 & 56613 & 87 & 2586 & 1.3 & 2.4 & 310 & 180 \\
J1847$-$0427 & 18:47:18.86(5) & $-$04:27:59(2) & 28.49 & $-$1.12 & 259.24664288(2) & 5.9(5)  & 188.3 & 4.7 & 56341$-$56882 & 56611 & 71 & 9527 & 3.3 & 69 & 13 & 140 \\
  \hline \label{tab:newPSR}
 \end{tabular}
}
\vspace{-0.2\skip\footins}
 \begin{flushleft}
 $^{*}$ Note that PSR~J1835$-$0924b is unrelated to the previously known PSR~J1835$-$0924, which has a catelogue spin period of 859.192\,ms at a DM of 471\,cm$^{-3}$\,pc.\\
 $^{\dagger}$ For PSR~J1829$-$1011, the period derivative-related parameters are derived with the 2$\sigma$ upper limit of $\dot{P} <7.4\times10^{-17}$. \\
 $^{\ddagger}$ The reduced $\chi^{2}$ stated here represents the value before the application of EFAC. Note that the rest of the timing solutions have EFACs incorporated, bringing the reduced $\chi^{2}$ to unity.\\ 
 \end{flushleft}

\end{table}
\end{landscape}

Despite the improved acceleration search algorithm used in this analysis, no previously unknown relativistic binary pulsar has yet been found. However, all known relativistic binary pulsars in the survey region of the \PROCESSED{} processed observations have been re-detected with a higher significance than obtained in previous analyses (refer to Table~\ref{tab:knownbinary} in the Appendix for details).

\begin{table*}
  \begin{minipage}{18cm}
    \centering
    \caption{\textsc{tempo2} best-fit parameters for the \NEWSOLUTIONbin{} newly-discovered binary systems. PSR~J1101$-$6424 can be compared to that of PSR~J1614$-$2230 listed at the last column (data taken from \protect\citet{Demorest2010}). Values in parentheses are the nominal 1-$\sigma$ uncertainties in the last digits. The last panel shows derived parameters, the respective equations for which can be found in \protect\citet{Handbook2004}, except for the DM distance which is derived according to \protect\citet{NE2001model}.}
    \begin{tabular}{p{5.8cm}llll}
  \hline
  Parameter                                   & J1755$-$25    & J1244$-$6359     & J1101$-$6424       & J1614$-$2230 \protect\citep{Demorest2010}) \\
  \hline
  Right ascension, $\alpha$ (J2000)            & 17:55.6(5)   & 12:44:47.693(18) & 11:01:37.1923(5)   & 16:14:36.5051(5) \\
  Declination, $\delta$ (J2000)                & $-$25:53(7)  & $-$63:59:47.4(3) & $-$64:24:39.332(2) & $-$22:30:31.081(7) \\
  Galactic longitude, $l$ $(\degr)$            & 3.76         & 302.20           & 291.42             & 352.64  \\            
  Galactic latitude, $b$ $(\degr)$             & $-$0.34      & $-$1.13          & $-$4.02            & 20.19  \\ 
  Spin period, $P$ (ms)                        &315.19598238(5)& 147.27431048(2)  & 5.109272904279(2)  & 3.1508076534271(6)  \\
  Period derivative, $\dot{P}$ $(10^{-18})$    & $-\,^{c}$    & 4.4(8)           & 0.0018(2)          & 0.0096216(9)  \\
  Dispersion measure, DM $\rm(cm^{-3}\rm\,pc)$ & 751(3)       & 286(1)           & 207                & 34.4865  \\ 
  \hline
  Orbital period, $P_{\rm{orb}}$ (d)           & 9.69633(2)   & 17.170748(2)     & 9.6117082(3)       & 8.6866194196(2)\\ 
  Projected semi-major axis, $x$ (lt-s)        & 12.2838(2)   & 24.0329(2)       & 14.02466(3)        & 11.2911975(2)   \\ 
  Epoch of ascending node, $T_{\rm{asc}}$ (MJD)& 56904.1259(6)$^{d}$ & 56513.0(3)$^{d}$ & 55689.00791(3)     & 52331.1701098(3)\\
  $e\sin\omega$, $\epsilon_{1}$ $(10^{-6})$    & $-\,^{d}$    & $-\,^{d}$        & 25.6(13)           & 0.11(3) \\
  $e\cos\omega$, $\epsilon_{2}$ $(10^{-6})$    & $-\,^{d}$    & $-\,^{d}$        & 2.0(12)            & $-$1.29(3)\\   
  Inferred eccentricity, $e$ $(10^{-6})$       & 0.08932(3)$^{d}$&0.000179(19)$^{d}$& 26.0(13)           & 1.30(4)    \\
  Longitude of periastron, $\omega$ $(\degr)$  & 129.65(2)    & 144(7)           & 85(2)              & $-$ \\
  Minimum companion mass$^{a}$, $m_{\rm{c,min}}$ $(M_{\sun})$ & 0.40 & 0.57      & 0.47           & $-$ \\ 
  Median companion mass$^{b}$, $m_{\rm{c,med}}$ $(M_{\sun})$ & 0.48 & 0.68       & 0.57            & 0.500(6)$^{f}$ \\ 
  \hline
  Binary model                                 & DD           & DD               & ELL1               & ELL1 \\
  First TOA (MJD)                              & 56901         & 56503            & 56401              & 52469 \\
  Last TOA (MJD)                               & 56942        & 56968            & 56967              & 55330 \\
  Timing epoch (MJD)                           & 56922        & 56222            & 56597              & 53600 \\
  Points in fit                                & 29           & 25               & 42                 & 2206   \\
  Weighted RMS residuals ($\upmu$s)              & 705          & 451              & 22                 & 1.1 \\
  Reduced $\chi^{2}\,^{e}$                     & 1.4          & 1.8              & 0.7                & $-$ \\
\hline
  DM distance (kpc)                            & 10.3         & 5.6              & 4.5                & 1.2 \\
  Characteristic age, $\tau_{\rm{c}}$ (Myr)    & $-\,^{c}$    & 520              & 44000              & 5200 \\
  Spin down energy loss rate, $\dot{E}$ $(10^{30}\rm\,erg\rm\, s^{-1})$   & $-\,^{c}$ & 56 & 540                & 12000\\
  Characteristic dipole surface magnetic field strength, $B_{\rm{surf}}$ ($10^{10}$\,G) & $-\,^{c}$ & 2.6 & 0.0098 & 0.018 \\ 
 \hline \label{tab:solutionBinary}
 \end{tabular}
\vspace{-0.5\skip\footins}
 \begin{flushleft}
 $^{a}$ $m_{\rm{c,min}}$ is calculated for an orbital inclination of $i=90^{\circ}$ and an assumed pulsar mass of $1.35\,M_{\sun}$. \\
 $^{b}$ $m_{\rm{c,med}}$ is calculated for an orbital inclination of $i=60^{\circ}$ and an assumed pulsar mass of $1.35\,M_{\sun}$. \\
 $^{c}$ For PSR~J1755$-$25, the time span of the data is only $\sim0.1\,$yr, which is not enough to resolve the degeneracy between the position and the spin down rate. Hence no period derivate-related parameters can be derived. \\ 
 $^{d}$ For PSRs~J1755$-$25 and J1244$-$6359 the DD model is used. We quote $\rm{T_{0}}$ instead of $\rm{T_{asc}}$. $e$ is directly fitted for and not inferred from the $\rm\epsilon$ parameters.  \\
 $^{e}$ The reduced $\chi^{2}$ stated here represents the value before the application of EFAC. Note that the rest of the timing solutions have EFACs incorporated, bringing the reduced $\chi^{2}$ to unity. \\
 $^{f}$ This companion mass is measured from the detected Shapiro delay. \\
 \end{flushleft}
 \end{minipage}
\end{table*}

\begin{table*}
\centering
  \caption{Discovery parameters of \NEWNOSOLUTION{} newly-discovered pulsars from the HTRU Galactic plane survey, which have not yet enough TOAs to produce a coherent timing solution. All of these pulsars have been assigned a temporary name containing only two digits of declination. We list the Galactic latitude ($l$) and longitude ($b$), the spin period ($P$) and the DM of these pulsars. Values in parentheses are the nominal 1$\sigma$ uncertainties in the last digits. The distances are derived according to \protect \citet{NE2001model}. }  \label{tab:timing2}
\begin{tabular}{lllR{0.8cm}R{0.8cm}llR{0.8cm}}
  \hline
 PSR~name & R.A. (J2000)  & Dec. (J2000)   & $l$ & $b$ & $P$ & DM & Dist  \\
  & $(^{\rm{h m}})$ & $(^\circ\,')$ & $(^{\circ})$ & $(^{\circ})$ & (ms) & (cm$^{-3}$pc) & (kpc) \\
  \hline
J1227$-$63 & 12:27.2(5) & $-$63:09(7) & 300.20 & $-$0.41 & 444.57796(6)    & 359(4) & 8.4 \\
J1255$-$62 & 12:55.3(5) & $-$62:48(7) & 303.37 & 0.06  & 182.91150816(3) & 713(4) & 13.2 \\
J1349$-$63 & 13:49.3(5) & $-$63:56(7) & 309.24 & $-$1.77 & 373.0340(1)     & 478(3) & 9.6 \\
J1532$-$56 & 15:32.3(5) & $-$56:32(7) & 324.04 & 0.01    & 522.97708597(13) & 282 & 4.3 \\
J1612$-$49 & 16:12.9(5) & $-$49:27(7) & 332.99 & 1.28    & 192.68718417(14) & 439(5) & 7.9 \\
J1612$-$55 & 16:12.1(5) & $-$55:09(7) & 328.99 & $-$2.78 & 846.907(3)     & 312(8)  & 5.7  \\
J1627$-$49 & 16:27.7(5) & $-$49:54(7) & 334.37 & $-$0.71 & 623.678(3)     & 594(16) & 6.8 \\
J1627$-$51 & 16:27.5(5) & $-$51:08(7) & 333.46 & $-$1.54 & 439.6840(5)    & 201(4) & 3.5 \\
J1634$-$49 & 16:34.5(5) & $-$49:52(7) & 335.13 & $-$1.49 & 684.93649236(2) & 649(12) & 8.8 \\
J1638$-$44 & 16:38.3(5) & $-$44:40(7) & 339.43 & 1.54    & 568.0566269(3) & 494(5) & 6.7 \\
J1658$-$47 & 16:58.4(5) & $-$47:12(7) & 339.74 & $-$2.77 & 369.3649(14)   & 533(5) & 11.3 \\
J1710$-$37 & 17:10.5(5) & $-$37:30(7) & 348.82 & 1.33    & 792.65922183(3) & 346(15) & 4.7  \\
J1718$-$41 & 17:18.2(5) & $-$41:07(7) & 346.74 & $-$1.98 & 548.4192(12)   & 354(7) & 5.4 \\
J1720$-$36 & 17:20.9(5) & $-$36:53(7) & 350.51 & 0.02    & 92.13212(3)     & 341(1)  & 4.2   \\
J1723$-$38 & 17:23.0(5) & $-$38:20(7) & 349.56 & $-$1.15 & 765.79581473(10) & 472(2) & 5.8  \\
J1730$-$34 & 17:30.1(5) & $-$34:48(7) & 353.28 & $-$0.36 & 99.82956(15)   & 628(5) & 7.0 \\
J1732$-$35 & 17:32.5(4) & $-$35:05(7) & 353.31 & $-$0.94 & 126.689998553(5)& 340(2) & 5.1 \\ 
J1741$-$34 & 17:41.9(5) & $-$34:19(7) & 355.00 & $-$2.15 & 875.137(2)      & 241(8) & 4.1 \\
J1743$-$35 & 17:43.1(5) & $-$35:32(7) & 354.09 & $-$3.00 & 569.980(9)      & 174(5) & 3.3  \\
J1746$-$27 & 17:46.0(5) & $-$27:51(7) & 0.97   &  0.49   & 487.52760873(8) & 422(9) & 5.2 \\
J1748$-$30 & 17:48.0(5) & $-$30:17(7) & 359.12 & $-$1.14 & 382.7347(6)     & 584(5) & 8.1 \\
J1756$-$25 & 17:56.7(5) & $-$25:28(7) & 4.24   & $-$0.34 & 855.986(2)      & 706(8) & 9.6 \\
J1759$-$24 & 17:59.4(5) & $-$24:02(7) & 5.79   & $-$0.16 & 1457.739(11)    & 772(14)  & 10.9 \\
J1819$-$17 & 18:19.5(5) & $-$17:05(7) & 14.15  & $-$0.90 & 2352.135(15)    & 67(23)  & 1.9 \\
J1830$-$10 & 18:30.1(5) & $-$10:39(7) & 21.04  & $-$0.17 & 245.260(3)      & 203(2) & 3.6 \\
J1834$-$09 & 18:34.6(5) & $-$09:15(7) & 22.79  & $-$0.51 & 512.3273(10)    & 529(7) & 6.4\\
J1836$-$11 & 18:36.1(5) & $-$11:17(7) & 21.15  & $-$1.77 & 496.1931(11)    & 326(8) & 5.2 \\
  \hline \label{tab:newPSR2}
 \end{tabular}
\end{table*}

\section{Individual pulsars of interest} \label{sec:individuals}

\subsection{PSR~J1101$-$6424, a Case-A Roche-lobe overflow cousin of PSR~J1614$-$2230} \label{sec:1101}
\begin{figure}
  \centering
  \includegraphics[width=3.2in]{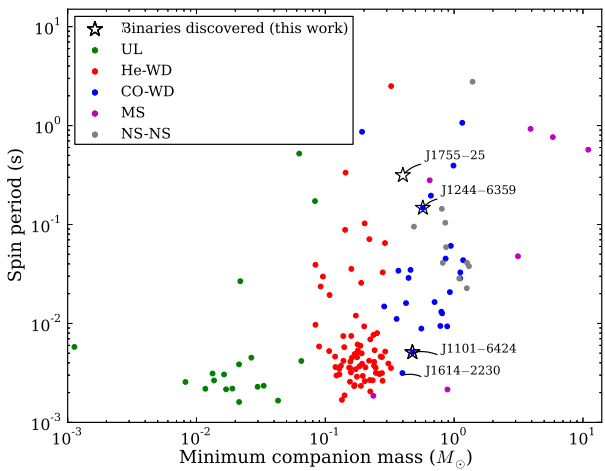}
\caption{Spin period versus minimum companion mass for all published pulsars in binary systems. We have classified their companions according to the description in \protect \citet{Tauris2012}. Green points indicate ultra-light (UL) binaries, red points indicate helium white dwarf (He-WD) companions, blue points indicate CO-WD, purple points indicate main-sequence star (MS) companion and grey points indicate NS-NS systems. The \NEWSOLUTIONbin{} newly-discovered binaries are additionally represented by star symbol, and only for PSR~J1755$-$25 the companion type is yet undetermined.}
\label{fig:J1101}
\end{figure}

PSR~J1101$-$6424 is a binary MSP with an orbital period of 9.6\,d and a spin period of 5.1\,ms (Table~\ref{tab:solutionBinary}). Assuming an orbital inclination of $i=90\degr$ and a pulsar mass of 1.35\,$M_{\sun}$, we find the minimum companion mass of PSR~J1101$-$6424 to be 0.47\,$M_{\sun}$. A heavy companion mass as such tends to point to an evolutionary track from an intermediate-mass X-ray binary \citep[IMXB;][]{Tauris2011}, with the companion being a carbon-oxygen white dwarf (CO-WD). However, the fast spin period of about 5\,ms indicates a rare example of full-recycling from a long mass-transfer phase. This can only be achieved via Case~A Roche lobe overflow (RLO) as discussed in \citet{Tauris2011b}, making PSR~J1101$-$6424 the second known system to descend from this IMXB evolutionary track. The only other binary MSP formed from Case~A RLO is an IMXB the 3.2-ms binary PSR~J1614$-$2230. Fig.~\ref{fig:J1101} plots the spin periods versus the minimum companion masses of all known pulsars as listed in \textsc{psrcat}. It can be seen that PSR~J1101$-$6424 is indeed located at the edge of the CO-WD population, together with PSR~J1614$-$2230. 

We recall that PSR~J1614$-$2230 is one of the heaviest neutron stars known, with a mass of 2.0\,$M_{\sun}$ \citep{Demorest2010}. This does not yet necessarily imply PSR~J1101$-$6424 should also have such a heavy pulsar mass. Nevertheless, the long-lasting accretion phase during Case A RLO leads to significant accretion onto the neutron star. Furthermore, given the similarities between the two systems (see Table~\ref{tab:solutionBinary}), we can deduce that the IMXB progenitor system of PSR~J1101$-$6424 is likely to have a similar initial donor mass of $\sim4.5\pm0.5\,M_{\sun}$, as estimated for PSR~J1614$-$2230 \citep{Tauris2011b,Lin2011}. According to Fig.~9 of \citet{Tauris2011b}, this would suggest a neutron star mass $>1.7\,M_{\sun}$ for PSR~J1101$-$6424. Any potential detection or constraint of Shapiro delay in PSR~J1101$-$6424 thus implies good prospects for measuring a pulsar with high mass, and continued monitoring of this binary system is of great interest. 

\subsection{PSR~J1244$-$6359, a mildly-recycled binary in a 17.2-d orbit }\label{sec:J1244}
This system is in a binary with orbital period of 17.2\,d and has a companion with a median mass of 0.69\,$M_{\odot}$ (Table~\ref{tab:solutionBinary}). The small orbital eccentricity of 0.000179(19) makes it unlikely that PSR~J1244$-$6359 is a DNS system. Rather, PSR~J1244$-$6359 is most probably a mildly-recycled binary system with a heavy CO-WD companion. The spin period of 147.3\,ms is on the slow side. However, Fig.~\ref{fig:J1101} shows it is not an outlier when comparing to the whole CO-WD binary population. The location of PSR~J1244$-$6359 on a $P$-$\dot{P}$~diagram is intriguing, bridging the population of normal and recycled pulsars (see Fig.~\ref{fig:PPdotHTRU}).

\subsection{PSR~J1755$-$25, an eccentric binary pulsar with a heavy companion} \label{sec:J1755}
PSR~J1755$-$25 was discovered with an unremarkable spin period of 315\,ms at a DM of 751\,cm$^{-3}$\,pc. We have now a coherent timing solution spanning over $\sim$40\,d and PSR~J1755$-$25 appears to be in a binary with an orbital period of 9.7\,d and an unusually high orbital eccentricity of 0.09 with a heavy companion of minimum mass $m_{\rm{c,min}}=0.40\,M_{\sun}$ (see Table~\ref{tab:solutionBinary}). Due to the short span of the timing data we cannot yet fit for the pulsar position and the period derivative. The covariance between these two parameters will be resolved when more timing data become available. The nature and the binary evolution scenario of this system is interesting and will be presented in a future publication.

\subsection{PSR~J1757$-$27, an isolated recycled pulsar} \label{sec:1757}
We have a coherent timing solution for PSR~J1757$-$27 across a time span of 557\,d with observations taken using the Lovell telescope. PSR~J1757$-$27 has a spin period of 17\,ms and a $\dot{P}$ of the order of 10$^{-19}$. The small $\dot{P}$ implies that PSR~J1757$-$27 is not a young pulsar. Rather, PSR~J1757$-$27 appears to be an isolated recycled pulsar, or in a very wide binary orbit with $P_{\rm{orb}}$ of the order of years. Further timing observations is crucial for revealing the nature of this pulsar. If it is indeed proved to be isolated, it will add to the currently small population of 30 Galactic (i.e., not associated with a Globular Cluster) millisecond-period isolated pulsars. This system would have interesting implications to the evolution scenarios of millisecond-period isolated pulsars. Currently, the most adopted formation scenario of isolated MSPs is that a pulsar system with a very-low mass companion could continue ablating its companion until it ceases to exist \citep[see e.g.,][]{Ruderman1989}. However, this would imply fully recycling and hence the pulsar should have very fast spin period of the order of a few ms, at odds with the 17\,ms spin of PSR~J1757$-$27. Another possible scenario is that PSR~J1757$-$27 comes from a disrupted binary system of Double Neutron Star (DNS). Given that PSR~J1757$-$27 appears to have been recycled, it would have to be the first formed pulsar of the DNS, which has subsequently been separated from its companion during the secondary supernova explosion.

PSR~J1757$-$27 has a small duty cycle of 2.8~per cent. If timing residuals with an RMS of the order of a few $\upmu$s can be achieved with future timing observations, PSR~J1757$-$27 can potentially be a good candidate for employment in a pulsar timing array.

\subsection{PSR~J1759$-$24, an eclipsing binary system with a long orbital period}
PSR~J1759$-$24 was first discovered from a survey observation taken at MJD\,55675.9 with strong, unambiguous pulsar characteristics. However, over 20 subsequent observations conducted both at the Parkes and the Lovell telescope between MJD\,56408 and 56675 spending over 10\,hr at the same position failed to re-detect the pulsar. PSR~J1759$-$24 was detected again since MJD\,56728 with both the Lovell and the Parkes telescopes and remains visible at the time of writing.

Given the high DM of 772(14)\,cm$^{-3}$\,pc of PSR~J1759$-$24, it is highly unlikely that scintillation could be responsible for the intermittent detections. A tentative timing solution indicates that PSR~J1759$-$24 could be in an eclipsing binary system with a long orbital period of the order of years (see Fig.~\ref{fig:J1759}). Similar examples include PSR~J1638$-$4725 which has a $P_{\rm{orb}}$ of 5.3\,yr and a $\sim20\,M_{\odot}$ MS companion, where the pulsar is undetectable in the radio wavelength for $\sim1$\,yr around periastron (Lyne et al., in prep); or PSR~B1259$-$63 which has a $P_{\rm{orb}}$ of 3.4\,yr and a $\sim10\,M_{\odot}$ MS companion, where the pulsar is undetectable during a 40-d eclipse behind the MS star \citep{Johnston1992a}. Further observations are needed to determine the nature of this pulsar and the cause for the intermittency.

\begin{figure}
  \centering
  \includegraphics[width=3.2in]{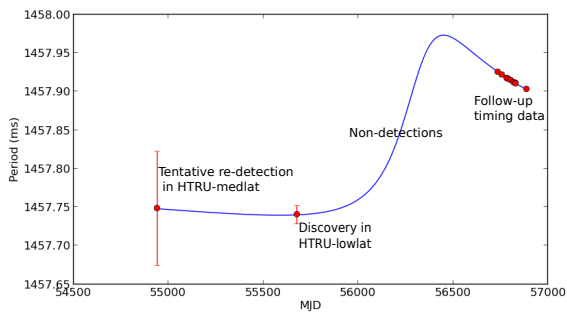}
\caption{A tentative orbital solution for PSR~J1759$-$24, with spin period plotted against the date of the observation in MJD.}
\label{fig:J1759}
\end{figure}

\subsection{PSR~J1847$-$0427, a pulsar with an extremely wide pulse} \label{sec:J1847}
PSR~J1847$-$0427 has a very wide profile with emission over almost the entire $360^{\circ}$ of pulse phase. Given that most radio pulsars have narrow emission patterns, PSR~J1847$-$0427 is likely an aligned rotator with an angle between the magnetic and rotational axis, $\alpha \sim0^{\circ}$. This would imply that we are sampling the emission pattern along a ring around the magnetic axis that is entirely within the beam of emission. On top of this, the profile of PSR~J1847$-$0427 exhibits a second component located at a rotation phase $\sim180^{\circ}$ from the main pulse (see relevant panel in Fig.~\ref{fig:profilenew}), similar to that observed in PSR~B0826$-$34 \citep{Biggs1988}. This second component could be due to oblateness in the emission beam. If future observations show evidences of drifting subpulses like the case of PSR~B0826$-$34, this could provide insight into the conical beam of emission. Further measurements of the polarisation properties and observations of any frequency evolution can help to solve for the emission geometry, and would also allow us to constrain the angle $\alpha$. These can potentially lead to studies of pulsar magnetosphere and testing of emission mechanism models.

\subsection{Two intermittent pulsars} \label{sec:intermittent}
PSRs~J1227$-$63 ($P=444.6$\,ms) and J1349$-$63 ($P=373.0$\,ms) were both discovered with constant radio emission over the entire course of the 72-min survey observations. Including the discovery, the subsequent confirmation and timing observations as well as a weak detection in PMPS archival data for the case of PSR~J1349$-$63, these two pulsars have been observed a total of 6.9 and 13.4\,hr respectively. However, they are not detectable in some of the follow-up observations,  

Given the high DM of 359(4) and 478(3)\,cm$^{-3}$\,pc of these two pulsars, it is highly unlikely that scintillation could be responsible for the non-detections. Instead, they are potentially nulling pulsars with long nulls lasting up to the $\sim$1-hr integration length of the follow-up observations, in which the pulsar is completely switched off \citep[see e.g.,][]{Kramer2006a}. Using Equation~(\ref{eq:SNExp}) and the observed pulsar duty cycle at discovery, we derive upper limits on the flux densities corresponding to the non-detections in the longest follow-up observations of PSRs~J1227$-$63 and J1349$-$63 to be 0.08 and 0.04\,mJy, respectively.

It is pre-mature to calculate the periodicities in the switching behaviour, but based on current data, we estimate that PSRs~J1227$-$63 and J1349$-$63 are respectively visible, on average, for less than 39 and 20 per cent of the time at 1.4\,GHz. Further observations are necessary to determine the nulling fraction of these intermittent pulsars.

\section{Comparing with known pulsar population} \label{sec:population}
\subsection{Luminosity}
\begin{figure}
  \centering
  \includegraphics[width=3.1in]{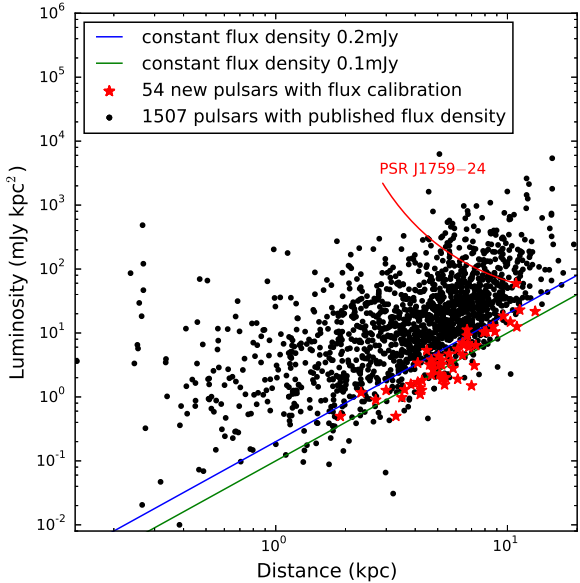}
  \caption{Luminosity versus distance of the \CALIB{} newly-discovered pulsars with flux density measurements (red stars) and the published pulsars taken from \textsc{psrcat} (black dots).}
\label{fig:L-dist}
\end{figure}

\CALIB{} out of the \NEWPSR{} newly-discovered pulsars have calibrated flux density measurements and we have inferred their luminosities as reported in column~8 of Table~\ref{tab:newSN}. We caution that the derived luminosities are dependent on our knowledge of the pulsar distances. In this work we have calculated pulsar distances based on the NE2001 electron density model \citep{NE2001model}, which is typically thought to have an associated uncertainty of 25~per cent for each DM-derived distance. Assuming that this uncertainty is of comparable magnitude irrespective of the line of sight, we compare our new luminosity derivations to the luminosity distribution of the known pulsars as listed on \textsc{psrcat}. It is interesting to see from Fig.~\ref{fig:L-dist} that the HTRU Galactic plane survey is indeed probing the low luminosity region particularly for distant pulsars, as expected from the improved frequency and time resolution as well as the longer integration time of 72\,min. One exception is PSR~J1759$-$24 which has the highest luminosity among the newly discovered pulsar. The intermittent detections associated with the eclipsing region have likely prevented this system (and possibly other intermittent pulsars along the Galactic plane) from being discovered previously. 

\subsection{Distance}
No nearby pulsars have been discovered thus far, out to a distance of roughly 2\,kpc (see Tables~\ref{tab:newPSR} and~\ref{tab:newPSR2}). This null result might seem surprising at first; plus a quick examination of Fig.~\ref{fig:L-dist} would suggest new discoveries to populate the region between the blue and green lines of constant flux densities. However, consider that the sampled sky volume increases with $D^{3}$, for a distance $D$. If we assume that pulsars are uniformly distributed in the Galaxy, there are thus more pulsars at larger distances. Furthermore, a uniform pulsar distribution is certainly not a good assumption. In fact towards the inner Galactic plane, the pulsar density should increase as our line-of-sight crosses more spiral arms and approaches the Galactic centre. This uneven Galactic distribution of pulsars further contributes to a higher number pulsars at larger distances for the HTRU Galactic plane survey region. Indeed, for the 713 known pulsars with a published luminosity within the HTRU Galactic plane survey region, only 27 (i.e., 3.8~per cent) lie within a distance of 2\,kpc. A Kolmogorov-Smirnov (KS) test comparing the newly-discovered pulsar distances with that of known pulsars within the same sky region of the HTRU Galactic plane survey indicates that the two samples are consistent with being drawn from the same population, with a p-value of 0.4 (i.e., probability > 0.05).

However, if the complete HTRU Galactic plane survey produces a significantly smaller percentage of nearby discoveries than 3.8~per cent, it could indicate that we have completed the nearby pulsar population, or at least reached a point where the yield of pulsar surveys are reducing as we are no longer flux limited. This has important implications to future pulsar surveys such as those to be conducted with MeerKAT and the SKA. Any pulsar surveys targeting the Galactic plane will have to go to higher observing frequency which could help with the discovery of pulsars that would be otherwise undetected due to scattering. A yet longer dwell time or a telescope with larger collecting area would thus be needed to compensate for the typically negative spectral index of pulsars as they get weaker at a higher observing frequency.

\subsection{Characteristic age}
Young pulsars (conventionally defined as having characteristic ages less than 100\,kyr) are expected to be located not far from their birth places hence mostly populating the Galactic plane region. Indeed the previous Galactic plane survey, the PMPS, discovered a sample of pulsars with an average characteristic age lower than the previously-known population \citep{PMPS2}, and these PMPS discoveries account for about half of the young pulsars currently known \citep{PMPS3}. No young pulsars have been discovered in the HTRU medium-latitude survey \citep{HTRU6}. This null result was attributed to the fact that the HTRU medium-latitude survey has a higher minimum detectable flux density of 0.2\,mJy compared to the PMPS of 0.15\,mJy along the Galactic plane. \citet{HTRU6} predicted that young pulsar discoveries are instead expected from the HTRU Galactic plane survey, given the higher sensitivity of this project part.

Of the \NEWPSR{} newly-discovered pulsars presented here, \HasPdot{} now have coherent timing solutions obtained over a time span of one year or more, allowing us to infer the preliminary characteristic ages and to locate them on a $P$-$\dot{P}$~diagram as in Fig.~\ref{fig:PPdotHTRU}. Note that we have not corrected for any potential contribution in the observed $\dot{P}$ from the Galactic potential and from the transverse proper motion of the pulsar \citep{Shk1970}. These effects are expected to be significant only for recycled pulsars \citep{Camilo1994}. 

\begin{figure}
  \centering
  \includegraphics[width=3.45in]{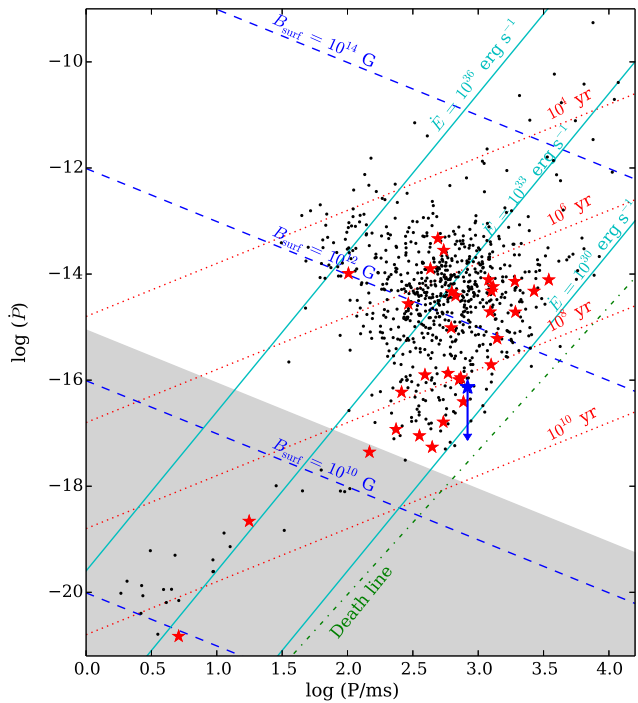}
\caption[Placing \PPDOT{} newly-discovered pulsars on a $P$-$\dot{P}$~diagram.]
        {$P$-$\dot{P}$~diagram showing the known pulsar population (black) within the HTRU Galactic plane survey region. The \PPDOT{} new discoveries that have now preliminary timing solutions are denoted by red stars, except PSR~J1829$-$1009 for which a 2$\sigma$ upper limit is shown by a blue star. Lines of constant characteristic ages are shown as red dotted lines, and it can be seen that the newly discovered pulsars appear to be of an older population. In addition, lines of constant spin-down energies are shwon as cyan solid lines and surface magnetic field strength as blue dashed lines. The death line as described in \protect\citet{Chen1993} is shown as green dot-dashed line. We define `recycled pulsars' to be those with $B_{\rm{surf}} \le 3\times10^{10}$\,G as indicated by the shaded region. } 
\label{fig:PPdotHTRU}
\end{figure}     

There is a noticeable lack of young pulsars amongst the \PPDOT{} which have timing solutions as they all have $\tau_{\rm{c}}$ at least of the order of Myr (also see Table~\ref{tab:timing1}). We caution that the inferred characteristic ages for recycled pulsars (shaded region in Fig.~\ref{fig:PPdotHTRU}) are known to be unreliable estimates in general \citep{Tauris2012a,Tauris2012}. Hence, we disregard all recycled pulsars, both previously-known and newly-discovered from this survey (namely PSRs~J1101$-$6424 and J1244$-$6309), and we compute a KS test comparing the ages of the newly-discovered pulsars with that of the known pulsars within the same sky region of the HTRU Galactic plane survey. The result indicates that the two samples are not consistent with being drawn from the same population, with a p-value of 0.005 (i.e., probability < 0.05). This is at odds with the expectation mentioned above. The identity of this older and less luminous population is intriguing, and may account for the missing population of disrupted mildly-recycled pulsars along the Galactic plane as predicted by \citet{Lorimer2004} and further investigated by \citet{Belczynski2010}, a population that has as yet largely eluded pulsar surveys due to their small average flux density. We note that the Galactic plane location of this old population also implies that these pulsars have likely received small kicks at birth.

In addition, there is a long standing debate regarding the relationship of radio luminosity of pulsars and their ages. In principle, the radio emission only contributes to a small portion of the energy budget of a pulsar. However, \citet{Narayan1990} suggested that radio luminosity is proportional to the cube root of the spin-down luminosity of a pulsar (i.e., $L \propto \dot{P}^{1/3}/P$). In other words, older pulsars would have a lower radio luminosity. \citet{Lorimer1993} showed that there is a large scatter in this relationship. Nonetheless, \citet{Arzoumanian2002} argued that the spread is simply caused by viewing geometry and intrinsically there is a strong luminosity-age relationship that scales with the voltage drop above the polar cap. If this is true, it would provide an explanation for the discovery of these older pulsars, as the HTRU Galactic plane survey probes deeper in the low luminosity end of the Galactic pulsar population.

Timing solutions for the remaining \NEWNOSOLUTION{} pulsars currently without a $\dot{P}$ measurement as well as any future discoveries from the HTRU Galactic plane survey will be crucial to differentiate if the current lack of young pulsars is purely due to small number statistics. The study of the age distribution of this less-luminous pulsar population probed by the HTRU Galactic plane survey should help to resolve some of the degeneracy of the arguments introduced above.

\section{A COMPARISON with the ESTIMATED DETECTIONS} \label{sec:yield}

Using the \textsc{psrpop}\footnote{http://psrpop.sourceforge.net} software based on the pulsar population model by \citet{Lorimer2006}, \citet{HTRU1} estimated that the HTRU Galactic plane survey should detect 957 normal pulsar (here defined to be of spin period longer than 30\,ms). We have conducted a similar simulation using the more up-to-date \textsc{PsrPopPy}\footnote{https://github.com/samb8s/PsrPopPy} software as presented in \citet{Bates2014}. By assuming the same power-law luminosity model as employed in \citet{HTRU1}, we find that the expected number of normal pulsar detections is 1020, statistically consistent with the result in \citet{HTRU1}.

Scaling these estimated numbers down to reflect the current HTRU Galactic plane survey data processing progress of \PROCESSED{}, it can be seen that of the order of 470 to 510 normal pulsar detections are predicted. We note that, strictly speaking, such intrapolation is inaccurate, as it assumes pulsars to be uniformly distributed in the survey region. This is not true; for example, a higher density of pulsars should be observed along the line of sights that cut through any spiral arms. However, as shown in Fig.~\ref{fig:skyprocessed}, this \PROCESSED{} of processed observations are drawn from random selection, hence this intrapolated estimation should hold. 

Among the \REDETECTIONp{} known pulsar re-detections, 427 are normal pulsars. Adding this to the 58 normal pulsars discovered thus far, we have a current normal pulsar detection rate of 485. The close match between the estimated and observed yield is satisfying, and would imply that our survey is performing as expected. In addition, the ability of these simulation tools to closely reproduce the observed yield could be used as an indication of a sound understanding of the underlying pulsar population parameters. Nonetheless, we note that a more mature judgment can be formed when the complete HTRU Galactic plane survey is analysed in the near future.

The detectability of the fastest-spinning MSPs (here defined to be of spin period $\le$ 30\,ms) is subjected to different pulsar-searching selection bias and is often considered separately. \textsc{psrpop} and \textsc{PsrPopPy} have predicted 51 and 43 MSP detections respectively for the complete HTRU Galactic plane survey. A dedicated study of MSP luminosity by \citet{Levin2013} suggested an even higher MSP yield of 68. Hence within the current \PROCESSED{} of processed observations, these simulations predict an MSP detection rate between 25 and 34. Among the \REDETECTIONp{} re-detected known pulsars, \REDETECTIONmsp{} are attributed to previously known MSPs, and two additional MSP detections come from the new discoveries of PSRs~J1101$-$6424 (Section~\ref{sec:1101}) and J1757$-$27 (Section~\ref{sec:1757}). These sum to a current total MSP yield of \MSPyield{}. It thus appears that we are roughly a factor of three short of MSP detections. 

Perhaps the most relevant explanation here, is that we have only processed the HTRU Galactic plane data in a partially-coherent manner. The segmentation scheme means that we have not make use of the full sensitivity achievable by the complete 4300\,s observations. In fact, running the MSP simulation again with \textsc{PsrPopPy} for each of the parallel searches of halved, quartered, and one-eighth segment length, the corresponding predicted MSP detections would be reduced to 27, 22, and 16 respectively, for the entire Galactic plane sub-survey. The current number of \MSPyield{} MSP detections in \PROCESSED{} of processed data then become at least consistent. As discussed in Section~\ref{sec:Segmentation}, coherently analysing the full 4300\,s observation with acceleration search is our priority in future data reprocessing. 

Apart from that, a major caveat in the estimation of MSP detection is that, our understanding of the underlying MSP distribution is still inadequate and hampered by the small number statistics of known Galactic MSPs. For this reason, MSP population simulation was not considered in \citet{Lorimer2006}. In addition, \citet{Lorimer2013} pointed out potential issues involved in the MSP period model devised by \citet{Cordes1997} which is used as an input parameter in \textsc{PsrPopPy}. \citet{Levin2013} found a steeper luminosity distribution and suggested a larger $z$-height for MSPs than what was used in the simulation by \citet{HTRU1}. These disagreements in the input parameters mean that we should treat any predictions from MSP population simulations with care. 
 
Despite these uncertainties in the MSP population model, several other reasons could come into play reducing our MSP yield. \citet{Eatough2013} have provided a detailed discussion on the possible causes of MSP non-detection. For instance, they suggested that the influence of unfavourable orbital phases where the acceleration was not constant, the emission beam having precessed out of our line of sight, intermittency, as well as human inspection error during pulsar candidate selection could have accounted for some of the missed MSPs. 

Furthermore, scattering broadening caused by the interstellar medium is particularly harmful for MSP detection. Any broadening of a time scale of ms already represents a significant portion of the spin period of an MSP, and the fact that MSPs tend to have larger duty cycles (see Table~\ref{tab:Smin}) makes them more susceptible to profile broadening. Scattering is not well quantified and if this effect is underestimated, could partially explain the apparent low yield of MSPs. We expect that the large sample of pulsars (both normal and MSPs) to be detected by the HTRU Galactic plane survey will provide further information for a better quantification of the extent of scattering. 

The presence of RFI could also have reduced our sensitivity which would have particularly affected the detection of low luminosity pulsars. However, as demonstrated by the close match of yield in normal pulsar detections, the HTRU Galactic plane survey does not seem to have any obvious general loss of sensitivity, hence RFI should not be a significant factor for the explanation of our lack of MSP detections.

\section{DISCUSSION and CONCLUSION} \label{sec:conclusion}
The HTRU survey is a blind pulsar survey of the southern sky with the 64-m Parkes radio telescope \citep{HTRU1} and a twin survey of the northern sky with the 100-m Effelsberg radio telescope \citep{Barr2013}. The HTRU surveys have benefited from recent advancements in technology and provide unprecedented time and frequency resolution, making the HTRU the first complete all-sky survey and the best pre-SKA survey. Thus far, over 170 newly-discovered pulsars have been reported from the HTRU survey with Parkes \citep{HTRU1,HTRU2,HTRU4,HTRU6,HTRU7,HTRU10,HTRU11}, including 33 MSPs. 

The HTRU low-latitude Galactic plane survey promises to provide the deepest large-scale search conducted thus far for the Galactic plane region, where the most relativistic binaries are expected to be found \citep{Belczynski2002}. In addition, this survey will represent a unique record of the Galactic plane with high yet uniform sensitivity, enabling an unbiased Galactic census to explore the true boundaries of pulsar phase space. The discoveries from the HTRU will also provide valuable knowledge of the Galactic pulsar population for the planning of survey strategies with the SKA.

Searching for pulsars in high resolution observations with long integration lengths is computationally intensive, and the depth to which the data can be explored is limited by the available computing resources. In order to improve the searching algorithm, we introduce two RFI mitigation techniques in the time and Fourier domain. The Fourier domain method has been shown to be effective and for some cases pulsars have been discovered only when this technique is applied. Furthermore, we present the implementation of a novel partially-coherent segmented acceleration search algorithm, based on the time-domain re-sampling method. This segmented search technique aims to increase our chances of discoveries of highly-accelerated relativistic short-orbit binary systems. We show that a $r_{\rm{orb}} \approx 0.1$ results in the highest effectiveness of the constant acceleration approximation, hence we split the long data sets of the HTRU Galactic plane survey into segments of various lengths and search for the binary systems using an acceleration range appropriate for the length of the segment. We push the maximum achievable acceleration value to 1200\,m\,s$^{-2}$. Within this range we can find a double-pulsar-like system deeper in the Galaxy with orbital periods as short as 1.5\,hr, and explore the parameter space occupied by expected pulsar-black hole systems. Of the order of $1.8\times10^{24}$ computational operations per data set are required by the search configuration adopted in this survey. The high computational requirements of such data-intensive astronomy also act as a test bed for SKA technologies. 

Analysis of \PROCESSED{} of the HTRU Galactic plane observations with the partially-coherent segmented acceleration search outlined above has resulted in \REDETECTIONu{} re-detections of \REDETECTIONp{} previously-known pulsars, demonstrating that the survey is performing as expected. In addition, we present the discovery of \NEWPSR{} pulsars, of which two are fast spinning pulsars with periods less than 30\,ms. One of the two pulsars, PSR~J1101$-$6424, is likely a descendant from an IMXB. Its fast spin period of 5.1\,ms indicates a rare example of full-recycling from a long mass-transfer phase, making it only the second known IMXB which has evolved from a Case~A Roche lobe overflow. The second one, PSR~J1757$-$27, is likely to be an isolated pulsar. The most adopted formation scenario of an isolated MSP requires the pulsar to have a very fast spin period of the order of a few ms, at odds with the unexpectedly long spin period of 17.7\,ms in the case of PSR~J1757$-$27. Furthermore, we report on the discovery of PSR~J1244$-$6359, a mildly-recycled 147.3\,ms pulsar in a 17.2-d binary orbit, bridging the normal and the recycled pulsar population on a $P$-$\dot{P}$~diagram. We report on the eccentric ($e=0.09$) binary pulsar PSR~J1755$-$25 with a heavy companion. We also report on the discovery of PSR~J1759$-$24. This pulsar is likely to be in an eclipsing binary system with a long orbital period of the order of tens of years. We report on the discovery of an aligned rotator PSR~J1847$-$0427 whose pulse profile contains a main pulse and a second component separated by 180$^{\circ}$, together forming a wide profile with emission over almost the entire 360$^{\circ}$ of longitude. We also report on the discovery of two intermittent pulsars, PSRs~J1227$-$63 and J1349$-$63. From the current discovery rate, extrapolation shows that the Galactic plane survey will result in roughly a further 60 discoveries. 
 
We note that the sky region of the HTRU Galactic plane survey has a complete overlap with that of the PMPS, as well as the HTRU medium-latitude survey. Despite the common sky coverage and the high success rates of these previous pulsar surveys, pulsar discoveries are still continuously being made from this survey. The improved dynamic range of the BPSR digital backend over the previous analogue filterbank system employed by the PMPS, as well as the long integration length of 72\,min of this survey, both account for the high survey sensitivity crucial for the discoveries of many less luminous pulsars. Nonetheless, some of the newly-discovered pulsars presented in this paper are later found to be detectable in the archival data. The discoveries of these relatively bright pulsars which have eluded previous survey efforts could be attributed to advancements in the data processing techniques. Particularly, the improvement in RFI mitigation techniques have reduced the number of false-positive candidates. Moreover, the `multiple-pass' nature of the partially-coherent segmented acceleration search has increased the chance of discovering pulsars by avoiding parts of the observation in which they might be less detectable. In principle, the acceleration search would also provide sensitivity to highly-accelerated relativistic pulsar binaries. However, none of the newly-discovered pulsars reported here was found at an orbital phase with noticeable orbital acceleration. 

Of the newly-discovered pulsars presented here, \PPDOT{} now have coherent timing solutions, which allows us to infer their preliminary characteristic ages. There is a noticeable lack of young pulsars among these pulsars as they all have $\tau_{\rm{c}}$ at least of the order of Myr. Using a KS test to compare the characteristic ages of this newly-discovered population with that previously known, we show that they are inconsistent of being drawn from the same population. The old age and low $\dot{E}/d^{2}$ of these pulsars make Fermi associations unlikely. Timing solutions for these pulsars and any future discoveries from the HTRU Galactic plane survey will be key to study the characteristic age distribution of this less-luminous pulsar population and might help to resolve some of the long standing arguments regarding the relationship of radio luminosity of pulsars and their ages.

A comparison with the estimated survey yield shows that we currently have a close match between the estimated and observed yield of `normal' pulsars. This is satisfying and would imply that our survey is performing as expected. We appear to be roughly a factor three short of MSP detections, and in Section~\ref{sec:yield} we explore the potential causes of the missing MSPs. We point out that the most relevant explanation is the fact that we have only processed the HTRU Galactic plane data in a partially coherent manner. The segmentation scheme means that we have not make use of the full sensitivity of the survey. Coherently analysing the entire observation length with an appropriate acceleration range is thus our first priority in future data reprocessing.

\section{acknowledgements}

The Parkes Observatory is part of the Australia Telescope National Facility, which is funded by the Commonwealth of Australia for operation as a National Facility managed by CSIRO. A large amount of the crucial computing resources needed for the data processing work is supported by the Australian National Computational Infrastructure (NCI) high performance computing centre at The Australian National University (ANU) and the ARC Centre of Excellence for All-sky Astrophysics (CAASTRO), as well as the HYDRA computer cluster funded by the Science and Technology Facilities Council (STFC). The authors would like to thank Anthony Holloway for providing constant support on the use of the HYDRA facilities, as well as the CAASTRO help desk for their every time swift replies. We thank Sally Cooper and Robert Dickson of the University of Manchester for helping the logistics of tape changing. We also thank Johnathon Kocz for useful discussion on RFI mitigation techniques, Thomas Tauris for sharing his knowledge on pulsar evolution, Paulo Freire for teaching the art of solving binary pulsars, Gregory Desvignes for his advices on pulsar timing, for reviewing the paper and providing many constructive suggestions, Lucas Guillemot for checking Fermi associations and Pablo Torne for carefully reading the manuscripts. We thank our summer student April Liska who has contributed to three of the pulsar discoveries presented in this paper, as well as the thorough work on database cross-checking by her and our work shadowing student William Mccorkindale. CN was supported for this research through a stipend from the International Max Planck Research School (IMPRS) for Astronomy and Astrophysics at the Universities of Bonn and Cologne.

\bibliographystyle{mn2e}

\appendix
\section{Technical details of the segmented acceleration search pipeline} \label{app:pipeline}
\subsection{Determining the optimal $r_{\rm{orb}}$} \label{app:rorb}
To quantify the effectiveness of a constant acceleration approximation versus varying $r_{\rm{orb}}$, we employed two observations of the double pulsar system PSR~J0737$-$3039A as test data sets. The double pulsar is the most relativistic pulsar binary system known, with very high maximum orbital acceleration of the order of 250\,m\,s$^{-2}$ and a short $P_{\rm{orb}}$ of 2.45\,hr. The orbital eccentricity is reasonably small ($e = 0.088$), so it closely reproduces the simplest orbital motion; a sinusoidal $v(t)$ of a circular orbit. Two test observations of the double pulsar system were carried out at Parkes. One was collected with the same observational set-up as the HTRU Galactic plane survey, where $t_{\rm{samp}}=64\,\upmu{\rm{s}}$ (hereafter test data set~1), and the other is identical to the test data set presented in \citet{Eatough2013}, where $t_{\rm{samp}}=80\,\upmu{\rm{s}}$ (hereafter test data set~2). The two different $t_{\rm{samp}}$ allow different $r_{\rm{orb}}$ to be probed, as we segment the test data sets into increasingly shorter $t_{\rm{int}}$, each with $n_{\rm{FFT}}$ of 2$^{k}$ to allow for maximum computational efficiency of the FFTs, where $k = 25,24,23,22,21,20$. 

For each $r_{\rm{orb}}$, we conduct a series of acceleration searches on a subset of the observation with the relevant length, incrementing at every 50\,s across the observed orbital phase. In Figs.~\ref{fig:SNseg} and~\ref{fig:SNseg2}, we plot the detected spectral S/N\footnote{The spectral S/N plotted here is the best S/N from the summing of the five harmonics mentioned in Section~\ref{sec:Processing}.} of the double pulsar in black and the recovered orbital acceleration in other colours. Panels~(a) of both figures show the shortest $r_{\rm{orb}}$ for each respective test data set. From the red data points it can be seen that these short segments with small $r_{\rm{orb}}$ each contain only a tiny fraction of the orbital motion of the double pulsar. Hence the analyses are not very sensitive to the trial acceleration value, leading to large fluctuations in the recovered orbital acceleration which does not follow the predicted grey curve closely. Nonetheless, the double pulsar has been detected throughout the orbital phase with roughly consistent spectral S/N, an indication that the constant acceleration approximation has been equally effective irrespective of orbital phase. As the coherent segment length gets progressively longer, as shown from panels~(b) to~(d), the acceleration searches become more successful in recovering the predicted orbital acceleration and the detected S/N also improves by roughly $\sqrt[]{2}$ as expected by the radiometer equation. Panels~(e) represent yet longer values of $r_{\rm{orb}}$ which exceed~0.1. At phases~0, 1, and~0.5, where the orbital acceleration is closest to being constant (i.e., the acceleration derivative $\dot{a}\approx 0$), improvements in the detected S/N are still observed compared to the previous $r_{\rm{orb}}$. However, the increase is less than $\sqrt[]{2}$, reflecting that the $r_{\rm{orb}}$ is now becoming too large resulting in spectral smearing which in turn reduces the detected spectral S/N, making the constant acceleration approximation less effective. At phases~0.25 and~0.75, where the orbital acceleration is increasing the most (i.e., a significant $\dot{a}$), the detected S/N is worse than the shorter $r_{\rm{orb}}$. Finally, panels~(f) show the longest $r_{\rm{orb}}$ where the degradation due to spectral smearing out-weighs the gain in S/N due to longer coherent segments, producing lower S/Ns at all orbital phases. The drastic drop in S/N immediately away from the orbital phases where $\dot{a}\approx 0$ is particularly noticeable. These are in agreement with the results of \citet{Eatough2013}.

\begin{figure}
\centering
\includegraphics[width=3.45in]{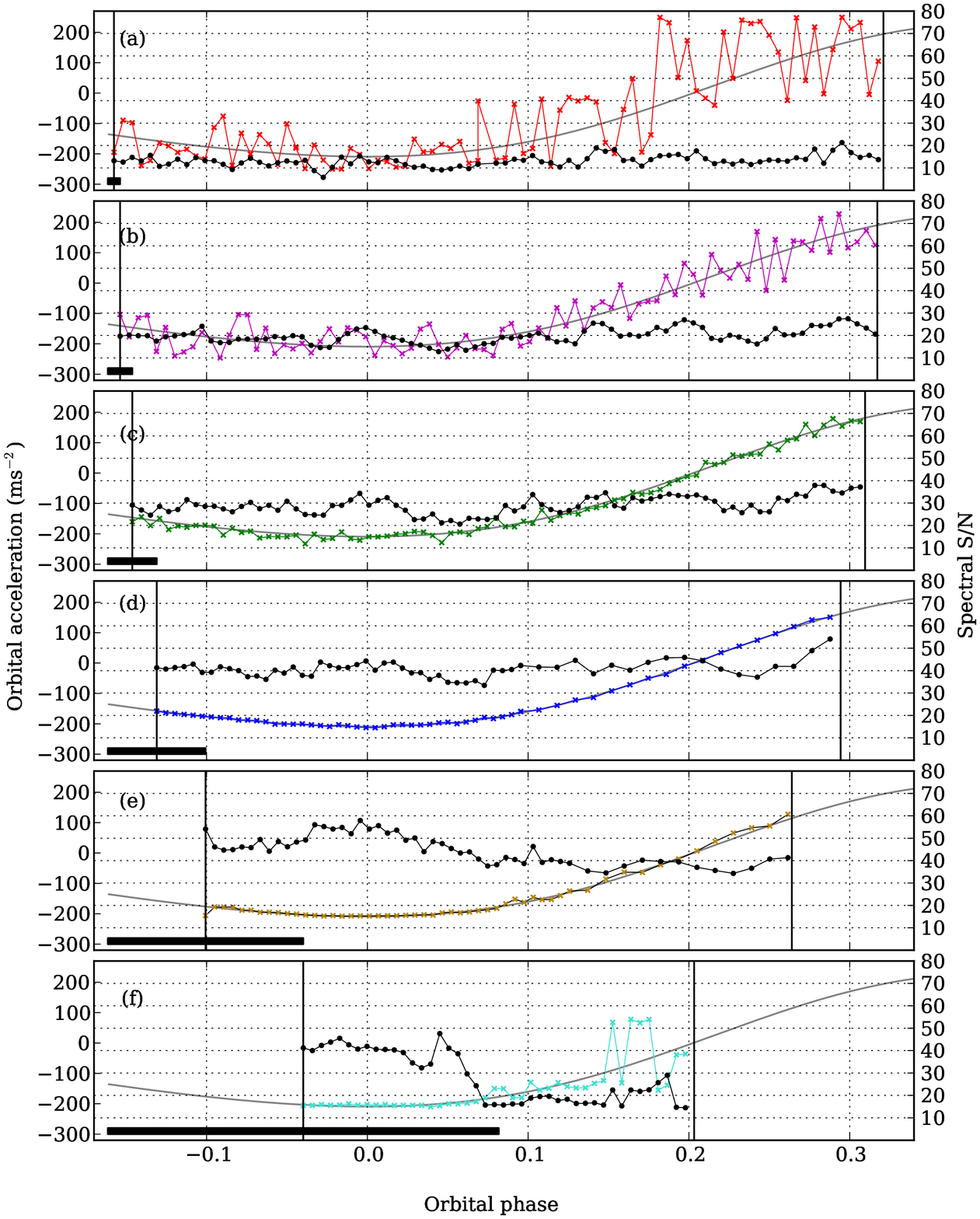} 
\caption{Detected orbital acceleration and S/N at various orbital phases for test data set~1. Each panel corresponds to progressively longer values of $r_{\rm{orb}}$, from (a)~0.0076, (b)~0.015, (c)~0.030, (d)~0.061, (e)~0.12 to (f)~0.24. The black bar on the bottom left of each panel depicts the length of each segment to be searched coherently. We slide this search window across the test data set incremented by every 50\,s, and we plot the acceleration search result at the middle of each segment. The recovered orbital acceleration values are plotted with colours, and the detected S/N is plotted in black. The smooth grey line shows the predicted acceleration curve.}
\label{fig:SNseg}
\end{figure}

\begin{figure}
\centering
\includegraphics[width=3.45in]{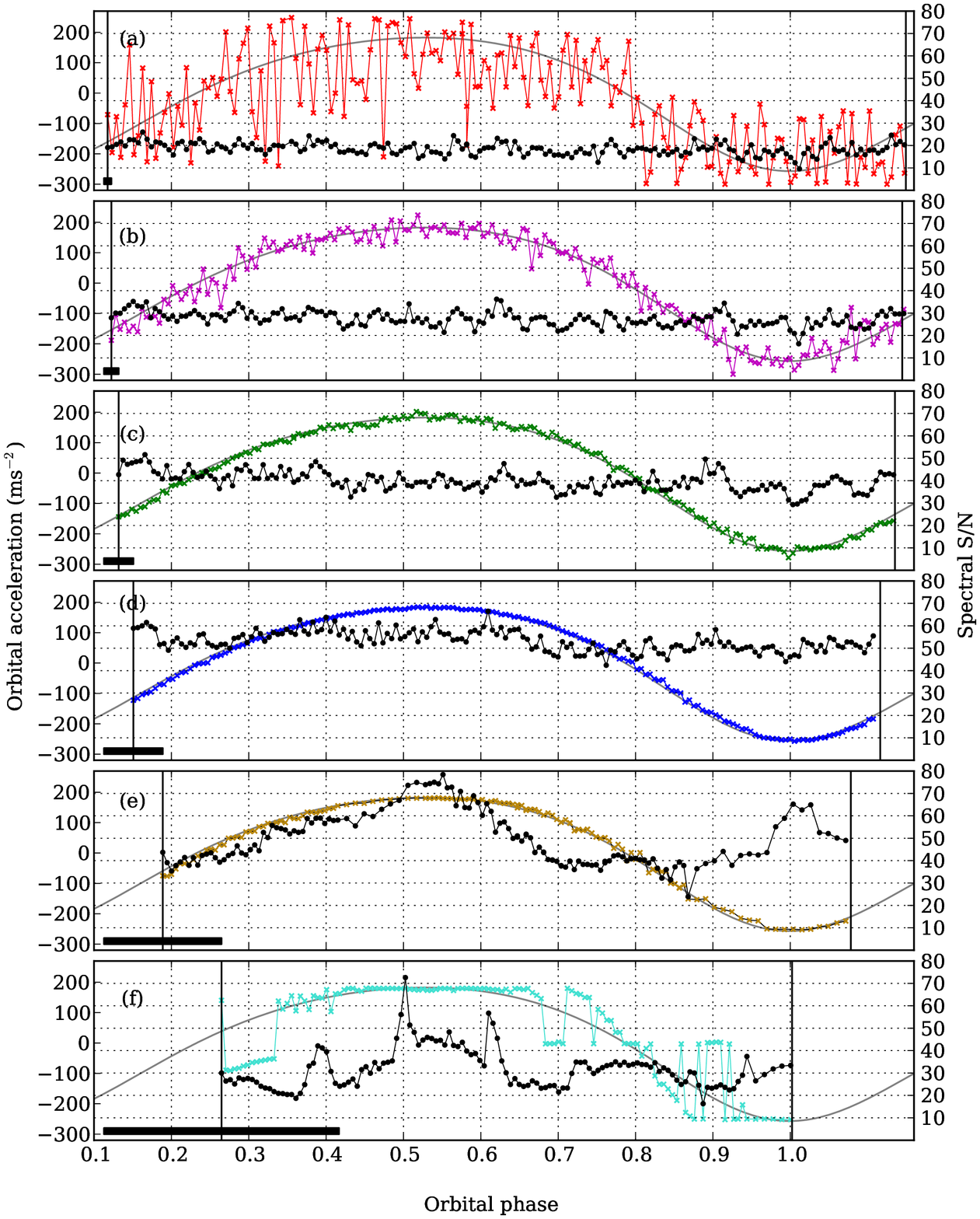}
\caption{Detected orbital acceleration and S/N at various orbital phases for test data set~2. Each panel corresponds to progressively longer values of $r_{\rm{orb}}$, from (a)~0.0095, (b)~0.019, (c)~0.038, (d)~0.076, (e)~0.15 to (f)~0.30.}
\label{fig:SNseg2}
\end{figure}

\subsection{Computational power considerations}
\begin{figure}
\centering
\includegraphics[width=3.3in]{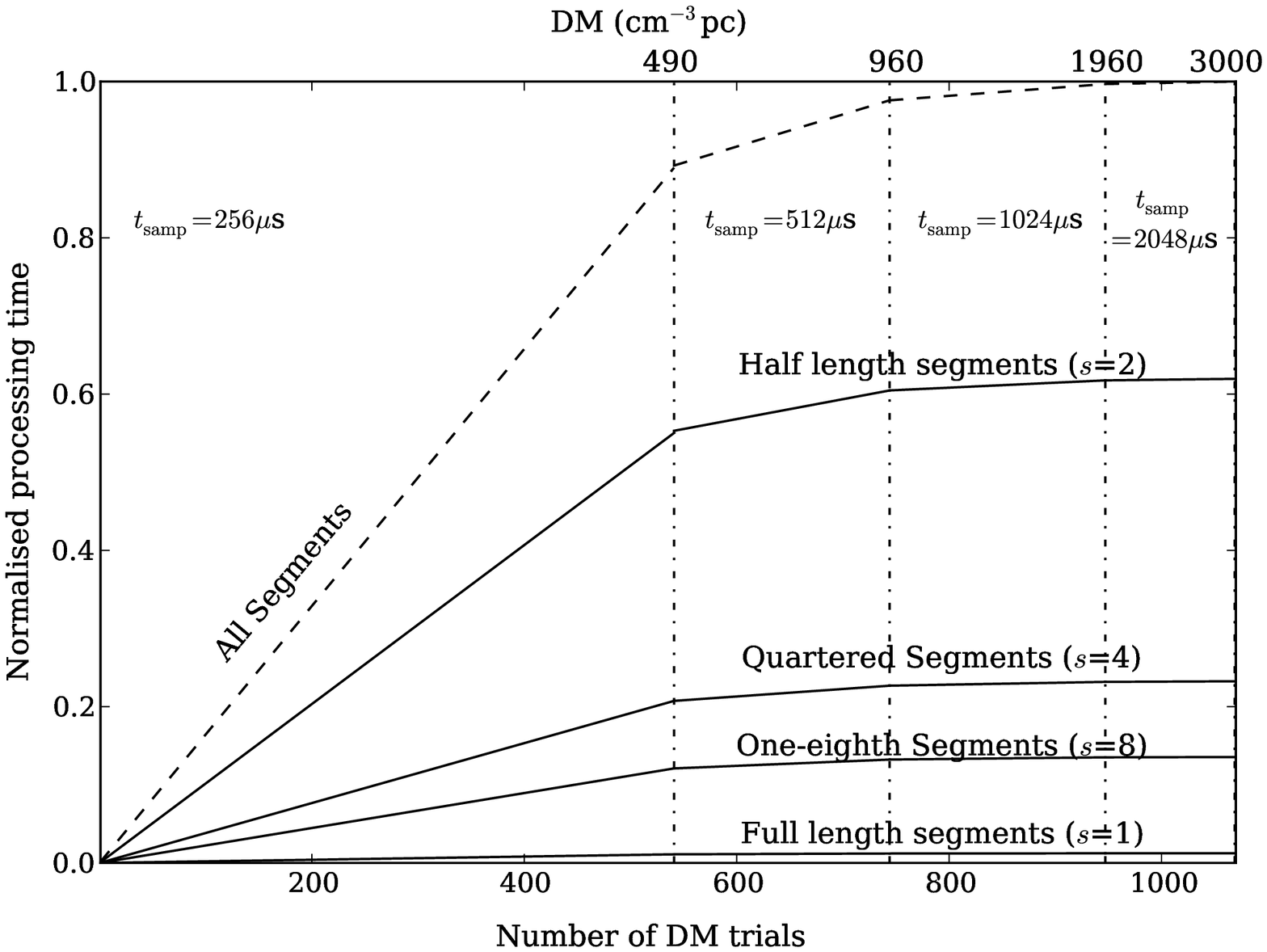}
\caption{The relative processing time required for each of the four configurations ($s=1,2,4,8$) parallel searches, normalised by the total processing time of all segments together. We plot these curves as a function of the number of increasing DM trials, and we show the corresponding DM values. The relevant $t_{\rm{samp}}$ is also marked.}
\label{fig:ProcessingTime}
\end{figure}

For this partially-coherent acceleration search, we adopt the same acceleration step size as described by Equation~(3) of \citet{Eatough2013}. \citet{Eatough2013} pointed out that pulsars with longer spin periods are less susceptible to a wrong acceleration trial (see Fig.~3 in their publication), and hence this acceleration step size is effectively slightly oversampling for these slow spinning binaries. In turn, the number of computational operations required for Fourier transforming each real time series, $C_{s}$, can be quantified as
\begin{equation}
C_{s} \propto \Delta{a} \times t_{\rm{samp}} \times {n_{\rm{samp}}}^{3} \times \log_2 ( n_{\rm{samp}})\,. 
\label{eq:C1}
\end{equation}
Note that because we progressively downsample in time after 2$\times$ the diagonal DM is reached \citep[see e.g., Section~6.1.1.2 of ][]{Handbook2004}, $t_{\rm{samp}}$ does not take the same value across different DM values, as indicated in Fig.~\ref{fig:flowchart}. Due to constraints of computational resources, we further downsample all observations to a $t_{\rm{samp}}\ge256\,\upmu$s. This might reduce our detectability towards MSPs with spin periods $\le1\,$ms, but still compares favourably to the $t_{\rm{samp}}=1\,$ms used in \citet{Eatough2013}. 

The total computational operations required to Fourier transform for each data set, $C_{\rm{tot}}$, then becomes
\begin{equation}
C_{\rm{tot}} \propto \sum\limits_{s=1,2,4,8}^S C_{s} \times {\rm{trials}}_{\rm{DM}} \times s\,,
\label{eq:Ctot}
\end{equation}
where trials$_{\rm{DM}}$ is the number of DM trials and its value is indicated in Fig.~\ref{fig:flowchart}. Summing the four configurations with segments of $s=1, 2, 4, 8$, we have of the order of $C_{\rm{tot}} \sim 1.8\times10^{24}$\,operations. Such processing typically requires $\sim$620\,CPU core hours on a single Intel Xeon Sandy Bridge node computer, for the analysis of one beam of an observation. Fig.~{\ref{fig:ProcessingTime}} illustrates the relative processing time needed for each of the four configurations as a function of the number of DM trials. 

\section{Candidate confirmation and gridding strategy} \label{sec:grid}
When a promising pulsar candidate is identified, it is necessary to conduct a confirmation observation at the telescope to verify if the candidate is a genuine pulsar. A successful re-detection confirms that the pulsar can be seen within the Gaussian beam of the receiver which has a FWHM of 14$'$.4 at Parkes. However, a better localisation of the pulsar position is desirable. Pulsars north of declination $-35\degr$ are followed-up with a timing campaign at the Lovell Telescope\footnote{In practice, for the declination between $-30\degr$ and $-35\degr$, only bright pulsars are followed-up at the Lovell Telescope. This is because of the short visible hours, as well as the challenges associated with such low elevation observations, namely the need for low wind conditions and the potential high spillover. (see Section~\ref{sec:new}), which has a smaller FWHM of 12$'$. In addition, an accurate position ensures efficient timing observations, as this maximises the S/N and hence reduces the telescope time required.}.

Major pulsar surveys at the Parkes telescope, such as the PMPS as well as the medium- and high-latitude parts of the HTRU, carried out a `N-E-S-W' gridding strategy as described in \citet{PMPS2} for achieving a better positioning of any newly-discovered pulsar. In addition to one re-observation at the exact position of the discovery (D), a grid surrounding the discovery position is carried out. An offset of $\delta_{1} =9\,'$ is applied in each direction of North (N), East (E), South (S), West (W) from the discovery position. This offset has been chosen so that these four grids form a tight square pattern through the discovery position (Fig.~\ref{fig:grid} upper panel). The Gaussian beam of the receiver would overlap enough between each grid position to ensure that the uncertainty of the pulsar position is less than a single beam width. The respective detected S/N from the `N-E-S-W' pointings can then be used to estimate the true position of the pulsar. 

\begin{figure}
\centering
\includegraphics[width=2.4in]{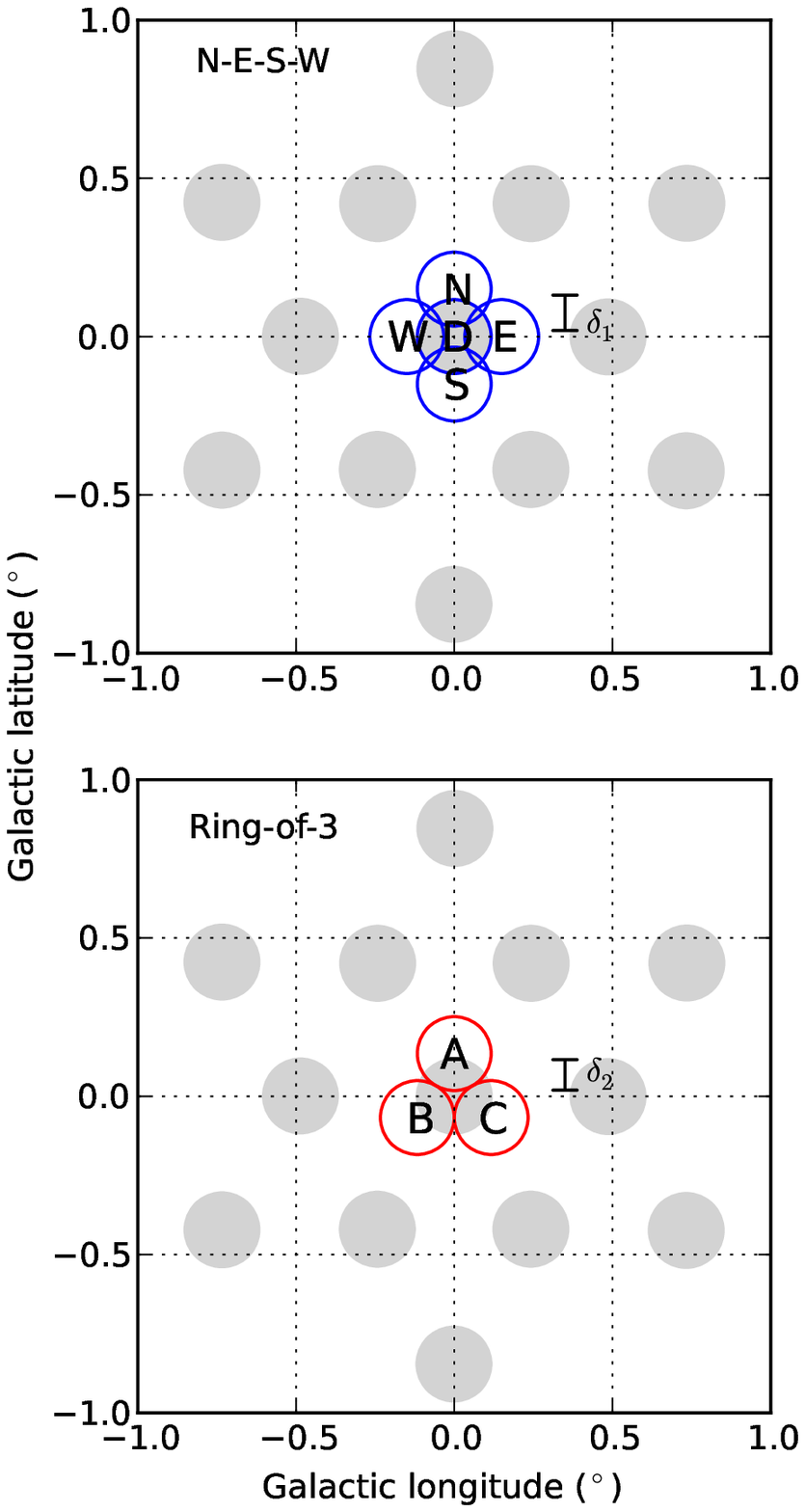}
\caption{Illustration of the gridding configuration. The light grey circles indicate beam pattern of the 13-beam Multibeam receiver with a FWHM of 14$'$.4. The upper panel shows the gridding strategy as described in \citet{PMPS2} where five positions (`N-E-S-W' and the discovery position `D') are required each offsetting the discovery position by $\delta_{1} =9\,'$. The lower panel shows the `Ring-of-3' configuration used in the gridding in this survey, where only three positions (A-B-C) are required each offsetting the discovery position by $\delta_{2} = 1/\sqrt[]{3}\times\rm{FWHM} \approx 8'$.3.} \label{fig:grid}
\end{figure}

The long integration length of the HTRU Galactic plane observations implies that, adopting this `N-E-S-W' gridding scheme as it is used in the other parts of the HTRU survey, a total of 72\,min\,$\times$\,5\,=\,6\,hr would be needed to confirm each pulsar candidate which is highly inefficient.  We hence employed an optimised strategy with a `Ring-of-3' grid (Fig.~\ref{fig:grid} lower panel) to minimise telescope time. Although this has a slightly worse sky coverage as compared to the `N-E-S-W' scheme, the factor of 5/3 gain in time is desirable. Furthermore, the integration time for each confirmation observation is scaled down from the nominal discovery S/N to achieve an expected S/N of 10 in the confirmation. Archival data (including overlapping observations from the HTRU medium-latitude survey and the PMPS observations) are checked before any gridding is carried out, in order to improve \textit{a priori} knowledge of the true position of the pulsar. In practice, we find that almost all of the \NEWPSR{} newly-discovered pulsars were successfully confirmed following this `Ring-of-3' scheme, affirming the efficiency over previous gridding strategies.

\section{PREVIOUSLY-KNOWN PULSARS WITHIN THE SURVEY REGION}
\subsection{A database of all re-detections} \label{app:redet}
In all, \REDETECTIONu{} re-detections of \REDETECTIONp{} previously known pulsars were observed in the HTRU Galactic plane survey data. Their details are listed in Table~\ref{tab:app} (full version available as Supporting Information with the online version of the paper).

\begin{table*}
\centering
\caption{The \REDETECTIONp{} previously known pulsars that have been re-detected in the \PROCESSED{} processed data of the HTRU Galactic plane survey. The file ID (which stands for the date and the time in UT of the observation) for which the pulsar was re-detected is listed, as well as the Galactic longitude ($l$) and latitude ($b$) that corresponds to the central position of the receiver beam of that observation. We list also the offset between this and the true position of the pulsar. Note that in the case when a pulsar was re-detected multiple times in different observations, we have listed only the re-detection which was closest to the pulsar, i.e. that with the smallest offset. We list the observed spin period ($P$) and the DM, as well as the expected flux density at 1.4\,GHz ($S_{\rm{exp}}$) and the expected S/N (S/N$_{\rm{exp}}$) calculated according to the radiometer equation. For the pulsars where no flux density is published on PSRCAT, we indicate with an asterisk (*). Finally, we list the observed S/N (S/N$_{\rm{obs}}$). The folllowing is a sample of the full table, which is available as Supporting Information with the online version of the article.} \label{tab:app}
 \begin{tabular}{p{1.4cm}p{3.0cm}R{0.5cm}@{}p{0.3cm}R{0.8cm}@{}p{0.7cm}lR{1.4cm}R{1.3cm}R{0.8cm}R{1.1cm}R{1.1cm}}
  \hline
PSR~name & file            & & $l$          & & $b$          &  offset      & $P_{\rm{obs}}$ & DM$_{\rm{obs}}$ & $S_{\rm{exp}}$ & S/N$_{\rm{exp}}$ & S/N$_{\rm{obs}}$ \\
         & Pointing/beam   & & $(^{\circ})$ & & $(^{\circ})$ & $(^{\circ})$ & (ms) & (cm$^{-3}$pc) &(mJy)    & \\
\hline
B0959$-$54 & 2011-10-11-18:36:49/01 & 280&.230 & 0&.060 & 0.03 & 1436.619$^{\ddagger}$ & 129.3 & 6.10 & 2100.0 & 1384.1 \\
B1011$-$58 & 2011-01-29-15:34:44/08 & 283&.587 & $-$2&.193 & 0.13 & 819.922 & 374.8 & 0.73 & 184.0 & 129.8 \\
B1014$-$53 & 2010-12-08-17:01:17/08 & 281&.170 & 2&.454 & 0.03 & 769.584 & 65.5 & 0.78 & 204.0 & 81.4 \\
B1015$-$56 & 2011-04-25-06:10:43/13 & 282&.712 & 0&.237 & 0.11 & 503.462 & 434.5 & 1.64 & 378.0 & 217.1 \\
B1030$-$58 & 2011-05-19-06:55:11/05 & 285&.963 & $-$0&.940 & 0.07 & 464.210 & 420.1 & 0.73 & 187.0 & 154.6 \\
B1036$-$58 & 2011-07-14-00:46:57/01 & 286&.300 & 0&.060 & 0.08 & 661.994 & 70.3 & 0.55 & 123.0 & 264.6 \\
B1039$-$55 & 2011-07-07-02:30:49/03 & 285&.235 & 3&.100 & 0.11 & 1170.865 & 299.4 & 0.34 & 75.5 & 160.2 \\
B1044$-$57 & 2013-02-25-12:20:14/12 & 287&.153 & 0&.683 & 0.10 & 369.431$^{\dagger\dagger}$ & 211.6 & 0.65 & 115.0 & 44.1 \\
B1046$-$58 & 2012-08-07-00:41:51/03 & 287&.335 & 0&.660 & 0.12 & 123.714 & 128.8 & 3.03 & 462.0 & 531.1 \\
B1054$-$62 & 2011-07-18-00:07:15/12 & 290&.413 & $-$2&.967 & 0.12 & 422.450 & 322.2 & 9.95 & 1250.0 & 1319.6 \\
\hline
\end{tabular}
\begin{flushleft}
 $^{\ddagger}$ Spin period originally detected at the third harmonics, i.e. at one third of the fundamental spin period listed here.\\
 $^{\dagger\dagger}$ Spin period originally detected at the half harmonics, i.e. at twice the fundamental spin period listed here.\\
\end{flushleft}
\end{table*}

\subsection{Non-detections} \label{app:missed}
Apart from the magnetar PSR~J1622$-$4950 discussed in Section~\ref{sec:known}, three further non-detections are related to a transient anomalous X-ray pulsar (AXP), probably in a radio quiet state. The non-detection of PSR~J1747$-$2958 can be explained by the high flux variability due to scintillation by the associated nebula. PSR~J1746$-$2849 is known to have a very broad profile and the fact that it is within $1^{\circ}$ from Sgr~A$^{*}$ explains its significant scattering time scale of $\tau_{\rm{sc,1.4\,GHz}}\,>\,226$\,ms, potentially preventing it from being detected in our survey. A further four non-detections are associated with pulsars with known intermittency and we might have observed them at a time when they were not emitting.

Excluding the magnetar, the AXP, the pulsars in local variable environment and the intermittent pulsars, twelve further non-detections are listed in the last three panels of Table~\ref{tab:missedPSR}. Five non-detections are attributed to the presence of RFI in individual survey observations or confusion with red noise. For PSRs~J1644$-$44 and J1644$-$46, the published declinations are not well-determined (E. Keane; private communication). This means that the actual S/N$_{\rm{exp}}$ might be lower if the true offset between the pulsar and the observed position is larger. The pulsars listed in the last panel of Table~\ref{tab:missedPSR} have unexpectedly weak S/Ns. Their relatively high DMs imply that scintillation is unlikely to have an influence on the detectability. It is not clear whether this is a result of an over-estimation in their catalogue flux densities or if some unknown phenomena such as nulling has occurred intrinsic to the pulsar. For PSR~J1746$-$2850, other observers have also failed in re-detecting any emission from other telescopes (G. Desvignes, P. Lazarus; private communication).

\begin{table*}
\begin{minipage}{18cm}
\setlength{\tabcolsep}{0.1cm}
\centering
 \caption{Previously known pulsars with an S/N$_{\rm{exp}} > 9$ that have been missed in the \PROCESSED{} processed observations of the HTRU Galactic plane survey. We list the file ID (which stands for the date and the time in UT) of the closest HTRU observation together with the positional offset from the catalogue pulsar position. The catalogue spin period and DM of the missed pulsars are also shown, together with the expected S/N (S/N$_{\rm{exp}}$) as well as the recovered S/N when folding observation with catalogue ephemerides (S/N$_{\rm{eph}}$).}
\begin{tabular}{llllp{0.9cm}p{0.7cm}p{0.7cm}p{7.3cm}}
  \hline
  PSR~name   & Observation ID &  offset       & $P$     & DM               & S/N$_{\rm{exp}}$  & S/N$_{\rm{eph}}$ & Comments\\
             & (Pointing/Beam)&  $(^{\circ})$ & (s)     & (cm$^{-3}$\,pc)  &           &  & \\
\hline
J1622$-$4950 & 2010-12-29-03:31:28/08 & 0.19  &  4.326100   & 820.0 & 20.2  & $-$  & Radio magnetar \citep{Levin2010}\\
J1809$-$1943 & 2011-05-17-17:28:45/10 & 0.21  &  5.540354   & 178.0 & 88.4  & $-$  & AXP XTE~J1810$-$197 \citep{Camilo2006} \\
             & 2011-06-26-12:10:21/03 & 0.22  &             &       & 78.1  & $-$  & \\
             & 2011-07-02-12:25:09/02 & 0.14  &             &       & 330.1 & $-$  & \\
J1747$-$2958 & 2013-01-07-03:05:18/08 & 0.12  &  0.098814   & 101.5 & 12.8  & $-$ & Associated with the `Mouse' radio nebula. Flux variability due to interstellar scintillation \citep{Camilo2002} \\
J1746$-$2849 & 2011-06-26-10:07:53/08 & 0.10  &  1.478480   & 1456.0 & 20.5    & $-$  & Very broad profile with $\tau_{\rm{sc,1.4\,GHz}}\,>\,226$\,ms and proximity to Sgr~A$^{*}$ \citep{Deneva2009}\\
\hline
J1634$-$5107 & 2011-05-07-19:53:44/13 & 0.07  &  0.507356   & 372.8 & 46.4  & $-$  & Known intermittency \citep{OBrien2006}   \\
J1726$-$31   & 2011-05-08-17:04:49/03 & 0.042 &  0.123470   & 264.4 & 12.2  & $-$  & Known intermittency, detection probability 20$\%$ \citep{Knispel2013}, published declination not well-determined.\\
J1808$-$1517 & 2013-04-02-18:57:44/10  & 0.13 &  0.544549   & 205.0 & 20.8  & $-$  & Known intermittency \citep{Eatough2013} \\
B1713$-$40   & 2013-04-01-14:21:48/05  & 0.23 & 0.887710    & 308.5 & 715.0 & $-$  & Known nulling. \citep{Wang2007} \\
\hline
B1727$-$33   & 2013-02-01-00:30:51/13 & 0.21 & 0.139460     & 259.0 & 36.6 & $-$  & Observation badly affected by RFI. \\
J1730$-$3353 & 2013-02-01-00:30:51/13 & 0.12 & 3.270242     & 256.0 & 38.4 & $-$  & Observation badly affected by RFI. \\
J1845$-$0316 & 2011-04-25-19:36:08/03 & 0.13 & 0.207636     & 500.0 & 20.9 & 13.2 & Observation affected by RFI. \\
J1840$-$0840 & 2013-01-06-02:28:11/12 & 0.21 & 5.309377     & 272.0 & 16.5 & 24.1 & Long period pulsar, confusion with red noise. \\
J1632$-$4621 & 2011-06-26-07:39:22/09 & 0.21 & 1.709154     & 562.9 & 29.8 & 15.2 & Relatively long period pulsar, confusion with red noise. \\
\hline
J1644$-$44   & 2011-07-03-10:59:47/11 & 0.047 & 0.173911    & 535.1 & 30.1 & 13.1& Published declination not well-determined. \citep{Knispel2013} \\
J1644$-$46   & 2011-05-17-13:02:58/02 & 0.044 &  0.250941   & 405.8 & 48.3 & 8.3 & Published declination not well-determined. \citep{Knispel2013} \\
\hline
J1749$-$2347 & 2013-02-02-01:44:13/08 & 0.067 &  0.874486   & 344.0 & 28.3  & 9.3  &  \\
J1809$-$2004 & 2013-04-02-16:32:16/05 & 0.18  &  0.434811   & 867.1 & 17.9  & 9.3  &  \\
J1818$-$1556 & 2013-04-07-16:12:18/03 & 0.15  &  0.952709   & 230.0 & 30.3  & 8.5  &  \\
J1746$-$2850 & 2011-06-26-10:07:53/08 & 0.12  &  1.077101   & 962.7 & 32.2  & $-$  & \\ 
J1644$-$4657 & 2011-05-17-13:02:58/08 & 0.098 &  0.125962   & 718.0 & 14.7  & $-$  & \\ 
  \hline \label{tab:missedPSR}
 \end{tabular}
\end{minipage}
\end{table*}

\subsection{Re-detections of previously-known binary pulsars}
Table~\ref{tab:knownbinary} lists all previously-known binary pulsars re-detected thus far. Two known binary systems (namely PSRs~J1525$-$5545 and J1807$-$2459A) which were undetected in our initial `standard search' pipeline were later found from the `partially-coherent segmented acceleration search' pipeline, and some other binaries have been detected with much higher S/Ns (for example in the case of PSR~J0737$-$3039A) from the acceleration search. These examples illustrate that the acceleration search algorithm allows us to detect fast binary systems which we would be insensitive to otherwise. 

\begin{table*}
  \begin{minipage}{18cm}
\setlength{\tabcolsep}{0.07cm}
\centering
 \caption{Previously-known binary pulsars re-detected in the \PROCESSED{} processed data of the HTRU Galactic plane survey, sorted by their respective catalogue orbital periods ($P_{\rm{orb,cat}}$) in descending order. We list the file ID (which stands for the date and the time in UT) of the relevant observation where these binaries were detected, as well as the detected spin period ($P_{\rm{obs}}$) and the DM of the pulsar. We compare the detected S/N from the `standard' (i.e., no acceleration search) pipeline with those from the full-length, half-length, quartered-length and one-eighth length segments of the `partially-coherent segmented acceleration search' pipeline described in Section~\ref{sec:Segmentation}. The corresponding detected orbital acceleration ($a_{\rm{orb}}$) from these acceleration searches are also listed. Note that in some cases, only one of the two pipelines has been employed, hence the non-applicable columns are denoted by `='. If the data were processed but the pulsar was not detected, it is denoted by `$\times$'.}
 \begin{tabular}{p{1.3cm}R{1.1cm}R{3cm}R{1cm}R{1.2cm}R{1.3cm}R{0.7cm}R{1cm}R{0.7cm}R{1cm}R{0.7cm}R{1cm}R{0.7cm}R{1cm}}
  \hline
   & & & & & `Standard' & \multicolumn{2}{c}{Full-length} & \multicolumn{2}{c}{Half-length} & \multicolumn{2}{c}{\nicefrac{1}{4}-length} & \multicolumn{2}{c}{\nicefrac{1}{8}-length} \\
  PSR~name & $P_{\rm{orb,cat}}$ & File ID & $P_{\rm{obs}}$ & DM$_{\rm{obs}}$ & S/N & S/N & $a_{\rm{orb}}$  & S/N & $a_{\rm{orb}}$ & S/N & $a_{\rm{orb}}$ & S/N & $a_{\rm{orb}}$ \\
           & (hr) & (Pointing/beam) & (ms)   & (cm$^{-3}$pc)  & & & (m\,s$^{-2}$) & & (m\,s$^{-2}$) & & (m\,s$^{-2}$) & & (m\,s$^{-2}$)\\
\hline
B1259$-$63    & 29681.4 & 2011-12-05-18:51:30/04 & 47.763  & 147.2  & 63.5 & 62.7 & 0.0 & 58.9 & 0.03 & 15.3 & 202.1 & 48.0 & $-$6.6 \\
J1711$-$4322  & 22139.3 & 2011-04-24-13:15:17/08 & 102.618 & 191.9  & 18.0 & 17.6 & 0.06 & 11.8 & $-$1.0 & $\times$ & $\times$ & $\times$ & $\times$  \\
B1800$-$27      & 9762.7 & 2011-08-16-12:50:12/12 & 334.412 & 162.1 & ==== &  48.1 & 0.0 & 42.3 & 1.0 & 24.1 & 206 & 28.7 & 104 \\ 
B1820$-$11    & 8586.3 & 2012-08-07-08:52:51/12 & 279.837 & 425.6   & ====  & 47.9 & 0.9   & 44.7 & $-$2.1 & 26.2 & 202.1 & 34.0 & 44.5 \\
J1708$-$3506  & 3579.2 & 2011-07-07-09:57:10/03 & 4.505   & 146.7   & 29.6 & \multicolumn{8}{c}{================================================} \\
J1750$-$2536  & 411.4 & 2013-02-06-20:15:56/10  & 34.745   & 179.2 & ==== & 11.5 & 0.064 & $\times$ & $\times$ & $\times$ & $\times$ & $\times$ & $\times$\\
J1810$-$2005  & 360.3 & 2013-04-08-20:07:00/12 & 32.822  & 238.8    & 9.1 & 28.3 & 0.33 & 25.4 & $-$1.0 & $\times$& $\times$& 18.2 & $-$6.6 \\
J1753$-$2240  & 327.3  & 2011-05-05-18:50:26/02 & 95.135  & 155.3    & 18.5 & \multicolumn{8}{c}{================================================} \\
J1454$-$5846  & 298.2   & 2013-01-02-19:24:06/05 & 45.249  & 116.8   & ====  & 25.5 & 0.06  & 19.9 & 1.0 &  $\times$&  $\times$&  $\times$&  $\times$   \\
J1543$-$5149  & 193.5  & 2011-07-04-07:23:31/07 & 2.057   & 51.1     & 14.8 & 12.9 & $-$0.2 & 9.7 & 0.03 &   $\times$&   $\times$&  $\times$&  $\times$  \\
J1232$-$6501  & 44.7  & 2011-04-23-08:08:01/01 & 88.282  & 240.0     & 36.5 & 39.3 & 0.9 & 31.6 & 1.0 & 8.9 & 308.5 & 17.4 & 27.4 \\
J1525$-$5545  & 23.8  & 2011-04-19-16:41:28/11 & 11.360   & 126.8     &  $\times$   & 24.0 & $-$1.0 & 32.4 & $-$1.0 &  $\times$&  $\times$& 17.7 & $-$6.6 \\
B1913$+$16$^{\dagger}$  & 7.8    & 2010-12-30-03:53:30/01 & 59.013 & 169.1  & 38.7 & 39.7 & $-$1.0 & 36.5 & $-$2.1 & 9.2 & $-$224.2 & 21.7 & $-$6.6 \\
J0737$-$3039A$^{\dagger}$ & 2.5 & 2010-12-30-17:11:35/01 & 22.674  & 47.9  & 14.1 & 14.1 & 0.0 & 32.0 & $-$198.9 & 38.2 & $-$207.1 & 38.0 & 163.6 \\
B1744$-$24A   & 1.8 & 2013-02-02-01:44:13/06 & 11.562 &  241.9 & ==== &  $\times$ &  $\times$ & 14.3 & 31.9 & $\times$&  $\times$&  $\times$&  $\times$   \\
J1807$-$2459A & 1.7 & 2011-05-08-15:24:15/04 & 3.059   & 134.3      &  $\times$    & $\times$    &  $\times$     & 19.2 & $-$3.2 & $\times$ & $\times$ & 14.2 & $-$6.6   \\
 \hline  \label{tab:knownbinary}
 \end{tabular}

\vspace{-0.7\skip\footins}
\begin{flushleft}
$^{\dagger}$ PSRs~B1913$+$16 and J0737$-$3039A are technically not within the HTRU Galactic plane survey region. However, these two binaries are highly relativistic and were observed as test pulsars at the beginning of the survey. The detected parameters are listed in this table, but they are not counted towards the \REDETECTIONp{} known pulsars re-detections. \\
 \end{flushleft}
 \end{minipage}
\end{table*}

\label{lastpage}
\end{document}